\documentclass[pre,showpacs,preprintnumbers,amsmath,amssymb]{revtex4-1}
\usepackage{graphicx}
\usepackage{bm}
\usepackage{color}
\usepackage{amssymb}
\usepackage{epsfig}
\DeclareGraphicsExtensions{.png,.pdf}

\begin{document}

\title{Spectral fluctuations of  multi-parametric complex matrix ensembles: evidence of a single parameter dependence}
\author{Mohd. Gayas Ansari and Pragya Shukla}
\affiliation{ Department of Physics, Indian Institute of Technology, Kharagpur-721302, West Bengal, India }
\date{\today}

\widetext

\begin{abstract}

We numerically analyze the spectral statistics of the  multiparametric Gaussian ensembles of complex matrices with zero mean and variances with different decay routes away from the diagonals.  As the latter mimics different degree of  effective sparsity among the matrix elements, such ensembles can serve as good models for a wide range of phase transitions e.g. localization to delocalization in non-Hermitian systems or Hermitian to non-Hermitian one.   Our analysis reveals a rich behavior hidden beneath the spectral statistics e.g.  a  crossover  of the spectral statistics from Poisson to Ginibre universality class  with changing variances  for finite matrix size, an abrupt transition for infinite matrix size and the role of complexity parameter,  a single functional of all system parameters, as a criteria to determine critical point.  We also confirm the theoretical predictions in \cite{psgs,  psnh}, regarding the universality of the spectral statistics in non-equilibrium regime of non-Hermitian systems characterized by the complexity parameter.

\end{abstract}

\def\stackalignment{l}

\maketitle

.

\section{Introduction}

Complex random matrices appear as the matrix representations of non-Hermitian operators of complex systems in diverse areas (e.g. see \cite{berry,  fh1, r9,  r7,  ahn, afm,  r10,  r5,  r8,kr1, prosen, hkku, bp, gnv, tk, r14, r15, r16, r20, r21,r22,ben1, ben0, ben2, fz, fy1, sc, fte1, fte2, fte3, fte4, cw, ns, cb, fy2, nh, fy3, haak, circ}).  
A knowledge of their statistical behavior is necessary to determine many of the physical properties.  As the  degree and type of randomness is often system specific and leads in general to a wide range of ensembles, it is  desirable to seek  a common mathematical formulation of their  statistical behavior; this would not only help to probe a specific complex system but also in revealing the connections hidden underneath among seemingly different complex systems.

The existence of a common mathematical formulation has been indicated in past for the Hermitian operators \cite{psall, psall1,  psrd, pscons2,  psrp,  psbe,  ptche, psijmp} and is also verified by the detailed numerical analysis on many body as well as disordered systems.  While a corresponding formulation for the  non-Hermitian cases was first derived in \cite{psnh} and further developed  recently in \cite{psgs} for the average spectral density, its detailed theoretical and numerical analysis on the complex plane e.g. the fluctuations of the radial and angular parts of the  eigenvalues, and validity of complexity parameter based predictions was not pursued earlier (mainly due to lack of interest in non-Hermitian operators in past). We note that contrary to Hermitian case, the statistical behavior on the complex plane  is far more richer, has many degrees of freedom and therefore technically far more complicated notwithstanding analogous basic mathematical skeletons in both the  cases. Intense renewed interest in the statistics of non-Hermitian operators but lack of a  theoretical formulation motivates us  to pursue, in the present work,  a detailed numerical analysis of the fluctuations.

As indicated by detailed theoretical, numerical as well as experimental studies,  the statistical behavior of complex systems with ergodic wave dynamics  can be well described by basis-invariant random matrix ensembles  i.e.  those with ensemble density dependent only on the  invariant properties of the matrix  e.g.  trace \cite{psijmp,  haak, gmw, me}.    In case of  non-ergodic wave dynamics,  however, the basis details affect the matrix representation of the complex system and lead to basis-non-invariant, multiparametric ensembles of random matrices.  As intuitively expected, the details of the ensemble are often sensitive to underlying system conditions but the relevant exact information is usually obscured by underlying complexity leaving access only to some statistical parameters. Fortunately,  based on the statistical approaches e.g. maximum entropy hypothesis, a knowledge of such parameters is often sufficient to predict the nature of the distribution e.g. Gaussian if the first two moments of the matrix elements are known.  This motivates us to consider a complex system, e.g a Hamiltonian with complicated many body interactions, represented by a complex matrix $H$ in an arbitrary orthonormal basis and 
with known constraints only on the mean and  variances of the matrix elements $H_{kl}$ and their pair wise correlations.
 Based on the maximum entropy hypothesis (MEH), it can then be described by an ensemble of $N\times N$ complex matrices $H$ defined by a Gaussian ensemble density 
\begin{eqnarray}
\tilde \rho(H,y,x) = C \; {\rm exp}\left[
- \sum_{s=1}^2 \sum_{k, l}
 (y_{kl;s} H_{kl;s}^2 + x_{kl;s} H_{kl;s} H_{lk;s})\right] 
\label{pdf}
\end{eqnarray}
 with $C$ as the normalization constant, $y$ and $x$ as the sets of  variances and covariances of various matrix elements, and, the subscript $s$ on a variable refers to  one of its components, i.e real ($s=1$) or imaginary ($s=2$).  
 
 For an ensemble to be an appropriate representation of the physical system of interest,  the ensemble parameters must be chosen as the functions of system parameters.  As the ensemble parameters $y_{kl;s}, x_{kl;s}$  in eq.(\ref{pdf}) can be arbitrarily chosen (including an infinite value for non-random entries),  this enables $\rho(H,y,x)$  to represent  a large class of non-Hermitian matrix ensembles e.g. varying degree of sparsity or bandedness.  Further by a variation of $y_{kl;s}, x_{kl;s}$, eq.(\ref{pdf})  can also be used as a model to analyze a wide range of crossovers/ transitions e.g. from Hermitian to non-Hermitian system conditions,  from one non-Hermitian universality class to another one e.g. Poisson to Ginibre universality class \cite{psgs, psnh}.

 The  standard route for the statistical analysis of a random matrix ensemble is based on the fluctuation measures of its eigenvalues and eigenfunctions and requires, in turn,  a knowledge of their joint probability distribution function (JPDF). An integration of the latter over undesired variables then, in principle, leads to the correlation functions for the  remaining variables e.g. the spectral correlations resulting from an integration over all eigenfunction components.  
 The integration is technically easier if the ensemble density is basis invariant e.g. depends only on the trace of the matrix \cite{psijmp}.  Except for the  basis invariant cases, an integration  of the ensemble density is in general technically complicated and a search for alternative routes is necessary.  As discussed in \cite{psgs, psnh},  a differential equation for the spectral JPDF  for non-Hermitian multiparametric Gaussian ensembles can be derived from the ensemble density.  
An integration of this equation over all  eigenvalues except one of them then leads to parametric evolution equation for the spectral density in the complex plane;  the solution of the equation under certain approximations and its verification by numerical analysis was discussed in \cite{psgs}.   But, contrary to the spectral density,  a derivation of its local fluctuation measures from  the sepctral JPDF  equation is  not an easy task and requires theoretical approximations.  (Indeed,  this is also the case for the  Hermitian ensembles: although the technical complexity here is relatively lower due to real  eigenvalues,   theoretical formulations are known either for second order spectral correlations for a few initial conditions or BBGKY hierarchical correlations among all orders \cite{psbe, fkpt5, haak, psijmp}).  

To gain an insight about the relevant theoretical approximation,  it is instructive  to first pursue a numerical investigation of the spectral fluctuations for the basis-dependent ensembles of complex matrices.  While the topic is rich in unknown vistas,  we confine our  focus in the present work on following objectives: 

(i) to establish whether the fluctuation measures satisfy local ergodicity assumption (needed to approximate spectral averages by the ensemble ones),

(ii) to analyze crossover of the fluctuation measures between Poisson and Ginibre universality class with varying ensemble parameters, 

(iii)  to analyze size-dependence of the  fluctuation measures and seek critical spectral statistics, 

(iv) to verify that different ensembles exhibit same spectral statistics if their complexity parameters are same and if they have same initial conditions.

The motivation for the first objective arises from the rich physics reported for non-Hermitian many body systems with non-ergodic dynamics. The ergodic dynamics is a necessary tool for system's approach to thermalization and thereby permitting application of  standard statistical tools e.g. equivalence assumption of spectral with ensemble averages.  While the spectral density for non-ergodic systems   is in general non-ergodic, this does not rule out local ergodicity of local fluctuation measures. and requires a numerical verification.  We note that the  ergodicity of the local fluctuations  mentioned above in (i) is different from the one discussed in \cite{psgs} in context of the average spectral density. 

The necessity for the second objective lies in the unavoidable presence of disorder as well as many body interactions in physical systems which can manifests in a variety of ways in the matrix representation of their non-Hermitian operators; it is natural to query whether presence of disorder with/ without  interactions  has any, and if so, what impact on the fluctuations of the excitations over representative ensemble and can these fluctuations mimic, at least locally, the fluctuations on a single spectral plane? .

A complex system often acts in an infinite dimensional Hilbert space and the random matrices in its representative ensemble are therefore of  infinite size. The numerical analysis however involves only finite size matrices. For application of  the numerical information to real systems, it is  necessary to analyze  how far the approximations and insights based on finite sizes be extended to  infinite sizes e.g.  due to some scaling relations  and  is the basis of our third objective. 

The quest for the fourth objective originates from the significant claim made in the study \cite{psnh}, indicating the existence of a single parameter governed common evolution equation for the spectral correlations for a wide range of physical systems represented by the Gaussian non-Hermitian random matrix ensembles. The system dependence in the formulation enters through the complexity parameter $Y$, a function of ensemble parameters and thereby a functional of system parameters \cite{psgs, psnh}. The common evolution equation predicts the analogy of the  statistical fluctuations for different non Hermitian Gaussian ensembles if their rescaled $Y$ values as well as the initial conditions are same.  This in turn claims an existence of universality in non-equilibrium regime of the ensemble representing a complex system (subjected to random perturbations); the significance of  the claim requires therefore a detailed numerical verification.

The paper is organized as follows. As mentioned above, the complexity parameter formulation of the spectral JPDF along with a detailed theoretical and numerical analysis of the average spectral density is discussed in \cite{psgs}. The present work pursues the study \cite{psgs} further, extending it to the local spectral fluctuations analysis on the complex plane. To  keep the present work self contained,  Section II briefly reviews the relevant steps of detailed theoretical formulation derived in \cite{psgs, psnh}.  Section III presents the ensemble densities  for the cases considered in our numerical analysis. 
To fulfil our objectives, here we numerically analyze  complex matrix ensembles for three different types of off-diagonal variances; the chosen ensembles are special cases of the ensemble density eq.(\ref{pdf}).  
A comparative analysis of the spectral fluctuation measures  for different ensembles requires a specific rescaling of the spectrum which in turn depends on whether the spectrum is ergodic or not; this is discussed in section III. A,  along with a numerical study of the ergodicity for the three ensembles along with the Ginibre ensembles.  Sections III.B and III.C present the  numerical analysis of the  fluctuations for each of the three ensembles for many system parameters; the objective of these sections is not only to verify that the crossover from initial to stationary state  is indeed driven by a single function of all system conditions but also to  provide insights about permissible approximations necessary for theoretical analysis of  mathematically intractable fluctuation measures.  Section III.D presents the details of a critical point analysis of the spectral statistics and reconfirm our theoretical predictions of \cite{psgs}. We conclude in section IV with a brief reviews of our main results and open questions.


\section{Complexity parametric formulation of Spectral correlations}

With ensemble (\ref{pdf}) representing a non Hermitian operator $H$ in an $N$-dimensional Hilbert space,  a typical  matrix $H \equiv \left[ H_{kl} \right]$ in the ensemble (\ref{pdf}) is a $N \times N$ complex matrix. The eigenvalues $\lambda_1,\ldots,  \lambda_N$ of $H$ are distributed on the complex plane with $\langle U_n |$ and $|V_n \rangle$ as the   left and right eigenvectors corresponding to $\lambda_n$ ($ \equiv \lambda_{n1}+i \; \lambda_{n2}$ with subscripts $1$ and $2$ referring to real and imaginary parts of the variable),
\begin{eqnarray}
\langle U_n | H = \lambda_n \langle U_n |, \hspace{0.8in}  H | V_n \rangle = \lambda_n  | V_n \rangle.
\label{big}
\end{eqnarray}
The eigenvectors are defined up to multiplication by scalar constants and are bi-orthogonal  i.e $\langle U_m | V_n \rangle = \delta_{mn}$.    (A nonhermitian matrix is considered to be  diagonalisable except at the exceptional points. The latter are the points in the parameter space at which some of the eigenvalues become degenerate and the corresponding eigenvectors coalesce into one.  But the number of  such  points is at most finite and it is  possible to avoide them by considering suitable  parametric conditions). 

The eigenvalue spectrum is described by the equation $H = U \Lambda V$ with  $\Lambda =\left[\lambda_n \delta_{mn} \right]$ as the eigenvalue matrix, $U \equiv \left[ U_{kn} \right]$, $V \equiv \left[ V_{kn} \right]$ as the left and right eigenvector matrices with $X_{kn} \equiv \langle k \mid X_n \rangle$ as the entries for $X= U, V$. 
Following from eq.(\ref{pdf}),  the  matrix elements  of  $H$ are randomly distributed. This in turn randomizes its eigenvalues and eigenfunctions thus rendering it imperative to consider their distributions.  Confining present analysis  to eigenvalues only,  the joint probability density function $\tilde {P}(z_1, z_2, \ldots z_N)$ (short notation for $P(z_1,  z_1^*,  \ldots z_N, z_N^*)$, equivalently $P(z_{11},  z_{12},  \ldots z_{N1}, z_{N2})$) for the eigenvalues  $\lambda_n $ to occur at $z_n$ on the spectral plane can be defined as 
\begin{eqnarray}
\tilde P(z_1, z_2, \ldots z_N) = \int  \prod_{s=1}^2 \;  \prod_{n=1}^N \delta(z_{ns}-\lambda_{ns})  \; \tilde \rho(H) \; {\rm D}H.
\label{pz}
\end{eqnarray}
with ${\rm D}H \equiv \prod_{kl;s}  {\rm d} H_{kl;s}$.

An integration of the spectral JPDF over $N-n$ variables $z_{n+1}, \ldots z_N$ leads to 
the probability densities $R_n (z_1,\ldots, z_n; Y)$ at each of the $n$ points $z_1,\ldots, z_n$ on the spectral plane irrespective of the rest $N-n$ variable positions.  Also referred as  $n^{\rm th}$ order spectral density correlator or simply n-level correlation functions,   can be defined as
 \begin{eqnarray}
 R_n(z_1,\ldots, z_n) = { N! \over {(N-n)!}}\int \tilde P(z_1,\ldots, z_N) \, {\rm  D}\Omega_{n+1}
\label{rn}
\end{eqnarray}
with ${\rm D} \Omega_{n+1}\equiv  \prod_{k=n+1}^N \; {\rm d} z_k \, {\rm d}z^*_{k}$.  The case $n=1$ corresponds to  the ensemble averaged spectral density $R_1$.   Alternatively,  assuming the  spectral density (defined as $\sigma(z) = \sum_{n=1}^N \delta(z-z_n) \delta(z^*-z_n^*)$)  consisting of  a secular part superimposed by a fluctuating part i.e $\sigma(z) = \langle \sigma(z) \rangle + \delta \sigma(z)$,  $R_n$  can also be defined as the correlation between $n$-point fluctuations of $\sigma(z)$:  $R_n(z_1,\ldots, z_n) = \langle \delta \sigma(z_1) \ldots \delta \sigma(z_n) \rangle$.    (The notation $\rho(z)$ used in the study \cite{psgs} for the spectral density  is replaced  here by $\sigma(z)$; this is  to avoid confusion with the notation $\rho(H)$ that is used in  {\it appendix } A).


All  spectral fluctuation measures can in principle be derived from the set of  $R_n$,   thus rendering the latter as the ideal tool for theoretical analysis.  But a determination of $R_n$ from eq.(\ref{rn}) requires a multi-dimensional integration besides a prior knowledge of JPDF $\tilde P(z_1, z_2, \ldots z_N)$.  Fortunately the technical complexity of the multi-dimensional integrals can be circumvented by a differential route \cite{psgs, psnh} that describes the response of the ensemble to varying system conditions (briefly discussed in {\it appendix A}).  As discussed in \cite{psgs, psnh},  the analysis leads to a common mathematical formulation of the evolution equation for $P(z_1, \ldots, z_N)$ and thereby $R_n(z_1,\ldots, z_n)$,  as system parameters  vary.   The evolution of both $P$ as well as $R_n$  on the spectral plane is however governed by a single functional $Y$ of the system parameters  (briefly discussed in {\it appendix A}) \cite{psgs, psnh} , 

\begin{eqnarray}
{\partial R_n \over\partial Y} &=&  \sum_{s=1}^2 \sum_{j=1}^n 
 {\partial\over \partial z_{js}} \left[ {\partial R_n\over \partial z_{js}} -  2 \,  { \partial {\rm ln} |\Delta_n(z)| \over \partial z_{js}} \, R_n
- 2 \, \int_{-\infty}^{\infty} {\rm d}^2 z_{n+1} R_{n+1} {\partial {\rm ln} |z_{j}-z_{n+1}| \over \partial z_{js}} \right]
\label{rnp1}
\end{eqnarray}
where $\Delta_n(z)$ is the Vandermonde determinant,

\begin{eqnarray}
\Delta_n (z) \equiv \prod_{k < l}^n  (z_k-z_l)
\label{delta}
\end{eqnarray}
and $Y$ is a function of all distribution parameters $x_{kl;s}, y_{kl;s}$.    For case $x_{kl;s} \not=0$,  we have \cite{psgs}

\begin{eqnarray}
Y = \mp {1\over  M} \sum_{k,l;s}  {\rm ln}\left(2(\gamma^2 + 2(-1)^s \tilde c_{kl;s}y_{kl;s}+ \gamma \sqrt{w})/y_{kl;s}\right)  + c_0.
\label{y} 
\end{eqnarray}
with  $y_{lk;s} = c_{kl;s} \; y_{kl;s}$,  $w={\gamma^2 + 4 y_{kl;s}(c_{kl;s} y_{kl;s} + (-1)^s \tilde c_{kl;s})}$ 
and the constants  $c_{kl;s}$ and $\tilde c_{kl;s}$ given by the relations    $x^2_{kl;s} + (-1)^s \; \gamma \; x_{kl;s} - c_{kl;s} \; y_{kl;s}^2 -(-1)^s\tilde c_{kl;s} \; y_{kl;s} =0$.  Here $M$ is the number of evolving parameters: $M=2 N^2$ for cases with $y_{kl;s} \not =0$.  (We note that, for the cases in which a specific $y_{kl;s}=0$ throughout the evolution, with $k, l, s$ arbitrary, the number $M$ of evolving parameter is less than $2 N^2$).  For the case $x_{kl;s}=0$,  we have $c_{kl;s}=0, \tilde c_{kl;s}=0$ and  therefore $w=\gamma^2$; $Y$ in this case is \cite{psgs, psnh}

\begin{eqnarray}
Y = \pm {1\over M}  \sum_{k,l;s}  {\rm ln} \; y_{kl;s}  + c_0,
\label{yt2} 
\end{eqnarray}

As $Y$ is a function of the ensemble parameters $y_{kl;s}$ and $x_{kl;s}$,  hereafter it will be referred as the ensemble complexity parameter.   We note that in case of  non-Hermitian complex systems (e.g. those with disorder or many body interactions) represented by the ensemble (\ref{pdf}),   $y_{kl;s}$ and $x_{kl;s}$ are in general functions of the system parameters e.g.  disorder, hopping,  interactions, dimensionality, boundary and topological conditions; as a consequence $Y$ is a functional of the system parameters.  As $Y$ appears as a "time" like parameter in the evolution described by eq.(\ref{rnp1}),  with $Y \to \infty$ corresponding to its stationary limit,  the solution of eq.(\ref{rnp1}) for finite $Y$ will be referred as the non-stationary or non-equilibrium states of the correlations. Although not relevant for the present study but worth recalling for completeness sake is the following aspect of the evolution of  $P$ and thereby $R_n$  in the complex plane: it is subjected to $M-1$ constants,  referred as $t_2,  \ldots, t_M$ in \cite{psgs}, which in turn depend on the matrix (global) constraints on $H$ and can  be chosen e.g. as the functions of the basis parameters (as the evolution of the ensemble density  occurs in a fixed basis-space in which the matrix $H$ is represented).

An additional  parameter,  although not present either in eq.(\ref{pdf}) or eq.(\ref{rn}) but appearing in eq.(\ref{y}),  is  $\gamma$; it is is a measure of the external confining potential,  the dynamics is subjected to 
(discussed in details in section II. A of \cite{psgs}).  But, for $R_n$ describing local spectral correlations in the bulk,  the confining potential does not affect them and therefore $\gamma$ does  not appear in eq.(\ref{rn}).   As the focus of the present study is an analysis of the bulk correlations (near $z \sim 0$), hereafter we set $\gamma=1$ in eq.(\ref{y}) without loss of generality.



As mentioned above,  $R_n$ for $n >1$  describe the  local fluctuations of the spectral density which is in general system-dependent.  Thus,  similar to Hermitian ensembles,  a  comparison of the local fluctuations imposed on different spectral density backgrounds  requires an "unfolding" of the levels.  The latter refers to  a  rescaling  of the eigenvalues $z_k$ by the local mean level spacing $\Delta(z)= (R_{1}(z))^{-1}$ at the spectral point of interest, say $z$, thereby resulting in a unit local mean level density in terms of the rescaled variables;  here $R_{1}(z)$ is the local mean level density of the eigenvalues  at $z$ before unfolding.   An important technical point here is the distribution of levels  on the complex plane,  each level  with two degree of freedom (real and imaginary parts).  In order to retain the same normalization of the mean level density before and after the rescaling,  it is important that both real and imaginary parts of $z_k$ should be rescaled by $\sqrt{\Delta(z)}$ . This follows from the conditions
 
 \begin{eqnarray}
{\mathcal R}_{1}(e)\; {\rm d}e \; {\rm d} e^* = R_{1}(z) \; {\rm d}z \; {\rm d}z^*,  \qquad {\mathcal R}_{1}(e) = 1.
\label{cond1}
\end{eqnarray}
The above conditions can be fulfilled e.g. by choosing 

 \begin{eqnarray}
e_{k1} = \int {\rm d}z_{k1} \; \sqrt{R_{1}(z)},  \qquad e_{k2} = \int {\rm d}z_{k2} \; \sqrt{R_{1}(z)}.  
\label{cond1}
\end{eqnarray}

%
%

The spectral correlations in terms of the rescaled complex variables $e_1, \ldots, e_n$ can be given as 

\begin{eqnarray}
{\mathcal R}_n(e_1,...,e_n; Y) =  {\rm Lim} N \rightarrow \infty \; {R_n(z_1,...z_n; Y) \over R_1(z_1; Y)....R_1(z_n; Y)} 
\label{rnu}
\end{eqnarray}


As $R_1(z_1) \approx R_1(z_2) \approx \ldots R_1(z_n)$ is assumed to be locally constant, 
we can approximate $\Psi \equiv  \prod_{k=1}^n R_1(z_k; Y) \approx \left(R_1(z; Y)\right)^n$ with $z$ as the spectral location where local density correlations are under consideration. 
The evolution equation for  ${\mathcal R}_n(e_1,...,e_n; Y)$ can now be obtained by
a substitution of $\Psi \;  {\mathcal R}_n(e_1,...,e_n; Y)$ on both sides  of  eq.(\ref{rnp1}). Subsequently the function $\Psi$ can be taken out of the derivatives with respect to $e_{ks}$ on the right side of eq.(\ref{rnp1}).  This leads to

\begin{eqnarray}
{\partial {\mathcal R}_n \over\partial \Lambda_e} &=&  \sum_{s=1}^2 \sum_{j=1}^n 
 {\partial\over \partial e_{js}} \left[ {\partial {\mathcal R}_n\over \partial e_{js}} -  2 \,  { \partial {\rm ln} |\Delta_n| \over \partial e_{js}} \, {\mathcal R}_n
- 2 \, \int_{-\infty}^{\infty} {\rm d}^2 e_{n+1} {\mathcal R}_{n+1} {\partial {\rm ln} |e_{j}-e_{n+1}| \over \partial e_{js}} \right]
\label{rnp2}
\end{eqnarray}
with $\Delta_n(e) \equiv \prod_{k < l}^n |e_k -e_l|$  and $\Lambda_e$ is  given by the 
relation  

\begin{eqnarray}
{\partial {\mathcal R}_n \over\partial \Lambda_e}  =  {1\over R_1 \, \Psi}  {\partial \Psi  {\mathcal R}_n \over\partial Y} =  {1\over R_1 }  {\partial {\mathcal R}_n \over\partial Y} +{{\mathcal R}_n \over R_1 }  {\partial \log\Psi \over\partial Y}
\label{rnp3}
\end{eqnarray}
Assuming  $R_1$ and thereby $\Psi$ to be varying slowly with respect to $Y$,  the term ${\partial \log \Psi \,  \over\partial Y}$  can be neglected,  leading to

\begin{eqnarray}
\Lambda_{e} = (Y-Y_0) \, R_{1}(z, Y).
\label{alme0}
\end{eqnarray}



As discussed in \cite{psgs},  however,   a $Y$-dependence of $R_1$ is sensitive to  the choice of the initial condition and may be too significant  to be  ignored; for example,  for a choice of a Poisson initial condtion, $R_1(z)$ varies rapidly for small $(Y-Y_0)$.    For such cases,  while the evolution of ${\mathcal R}_n$ can still be described by eq.(\ref{rnp2}) but eq.(\ref{alme0}) is no longer applicable for $\Lambda_e$.
%
%
This motivates us to define $\Lambda_{e}$  as 
\begin{eqnarray}
\Lambda_{e} = (Y-Y_0) \, R_{1, local}(z; Y).
\label{alme1}
\end{eqnarray}
where the function  $R_{1, local}(z; Y) $  is determined, in principle, from eq.(\ref{rnp3}); ( so named because the function  has the same units as the average spectral density).
Technically however this is not an easy task.  Indeed,  for similar studies on  Hermitian ensembles,  a more physically motivated, intuitive definition justified by numerical analysis has been used.  
 As indicated by these studies,  $R_{1,local}(e)$ for a  $d$-dimensional Hermitian Hamiltonian with matrix size $N$ depends on, besides $R_1(e)$,  on  the  eigenfunction correlations too (at energy $e$) and can be formulated as (e.g.  see section 6.13 of \cite{haak},  \cite{psijmp, psbe, psall, psrp} and references therein)
 \begin{eqnarray}
 R_{1,local}(e) = R_1 (e)\; {\eta(e) \over N}.
 \label{rloc}
 \end{eqnarray}
 Here  $\eta(e)$ is the average correlation/ localization volume at energy $e$ and is related to the inverse participation ratio $I_2(e)$ at  $e$: $\eta \propto (I_2)^{-1}$ for $e$ in the bulk of the spectrum.   The above definition indeed originates from the contribution  of each spectral level $e_n$,   weighted by its intensity $|U_n|^2$,  to local spectral density and can be defined as 
 $ R_{1,local}(e) = \langle \sum_{n=1}^N  |U_n|^2 \; \delta(e-e_n) \rangle$.  Using  $ |U_n|^2 \sim {\eta(e) \over N}$  then leads to  the definition in eq.(\ref{rloc}).   Intuitively it seems the same definition of  $R_{1,local}(e)$ can be extended to non-Hermitian case too (by replacing $|U_n|^2$ by $U_n^*V_n$).  Its theoretical verification however requires a statistical study of the eigenfunction correlations too which is beyond the ambit of present study.  To distinguish it from  $Y$  (the ensemble complexity parameter),  hereafter $\Lambda_e$ will be referred as the spectral complexity parameter.   





 We note that the rescaling of $Y$ leading to $\Lambda_e$ in the present case  is different from the Hermitian ensembles.  In the latter case, we have  $\Lambda_{e} = (Y-Y_0) \, R^2_{1}(e; Y)$  with $e$ as a real variable if $R_1(e; Y)$ is slowly evolving with $Y$  (e.g.  see section 6.13 of \cite{haak},  \cite{fkpt5} for Brownian ensembles),  and,   $\Lambda_{e} = (Y-Y_0) \, R^2_{1, local}(e)$ for cases in which $R_1(e; Y)$ varies significantly with $Y$ even at a local mean level spacing (e..g multiparametric Gaussian ensembles discussed in \cite{psijmp, psbe, psall, psrp} and references therein).   A different rescaling of $Y$ in non-Hermitian case arises from  the  rescaling of both real and imaginary  components of the eigenvalues.

The   rescaling of $Y$  introduces, in left side of eq.(\ref{rnp2}), an $e$-dependence (that was absent before) and  as a consequence,  the fluctuations of rescaled $n^{th}$ order correlations retain location-dependence even after unfolding (similar to the Hermitian cases discussed in \cite{psbe, ptche, psijmp, psall}); although the latter removes the dependence on the location $"z"$ from $R_n$  but reintroduces it through $\Lambda_e(z)$. We note that the original definition of unfolding was introduced in context of the transnationally-invariant, stationary spectrum  which corresponds to $\Lambda_e \to \infty$ limit of our formulation \cite{fkpt5}).  The location-dependence of ${\mathcal R}_n$  leaves them no longer translational as well rotational  invariant on the complex plane,  thus making it imperative  to analyze their radial and angular dependence.  To analyze the latter, it is more appropriate to first rewrite eq.(\ref{rnp1})  in polar coordinates. 
Writing  ${\mathcal R}_n(z_1, z_2,\ldots, z_n)$ as ${\mathcal R}_n (r_1,\ldots,r_n; \theta_1, \ldots,\theta_n)$ along with $e_k=r_k \; {\rm e}^{i\theta_k}$ and  ${\rm D} \Omega_{n+1}\equiv  \prod_{k=n+1}^N \; r_k {\rm d}r_{k} {\rm d}\theta_{k}$, 
eq.(\ref{rnp1}) can be rewritten as 

\begin{eqnarray}
{\partial {\mathcal R}_n\over\partial \Lambda_e}  &=&   \sum_k  \left({2\over r_k} {\partial  \over \partial r_k} + {\partial^2  \over \partial r_k^2} + {1\over r_k^2} \; {\partial^2  \over \partial \theta_k^2}\right) \, {\mathcal R}_n  + 2 \, \sum_{k,l; k \not=l} \left( {\partial  (C_{k l} \; {\mathcal R}_n )\over \partial r_k} + {1\over r_k} \, {\partial  (D_{k l} \; {\mathcal R}_n)\over \partial \theta_k}\right) - \nonumber \\
& & 2 \sum_k  \int {\rm d}r_{n+1} \; {\rm d}\theta_{n+1} \; r_{n+1}   \; \left[{ \partial  C_{k n+1} \over \partial r_k} + {1\over r_k} \, {\partial  D_{k n+1} \over \partial \theta_k}\right] \; {\mathcal R}_{n+1} 
\label{rna3}
\end{eqnarray}
where $C_{km}(r_k, \theta_k, r_m, \theta_m) = { r_k- r_m \; \cos(\theta_k-\theta_m)  \over r_k^2 + r_m^2 -2 \, r_k \, r_m \cos(\theta_k -\theta_m)}$, 
$D_{km}(r_k, \theta_k, r_m, \theta_m)  =  { r_m \, \sin(\theta_k-\theta_m)\over r_k^2 + r_m^2 -2 \, r_k \, r_m \cos(\theta_k -\theta_m)}$.

The above equation  describes evolution of ${\mathcal R}_n$ on $r, \theta$-plane from an arbitrary initial state  at   $\Lambda_e=0$ (equivalently $Y=Y_0$) to the stationary  limit at $\Lambda_e=\infty$ (equivalently $Y \to \infty$).  
Its general solution gives in principle the radial and angular dependence of ${\mathcal R}_n$ for arbitrary $\Lambda_e$ as well as the initial conditions.  In many cases, such multivariable equations can approximately be solved by a separation of variables approach but strong correlations rule out that possibility in the present case.  As the theoretical determination of the solution is technically very complicated, it would be helpful  to first develop some insights e.g by alternative routes.   This motivates us to consider, in the next section,  a numerical analysis of the spectral fluctuation measures which are not only  numerically and experimentally accessible but also depend on  many orders of the correlations ${\mathcal R}_n$.   

{\it Universality and criticality of the fluctuations:}    With ensemble dependence in eq.(\ref{rna3}) appearing only through $\Lambda_e$ (besides constants of evolution),  ${\mathcal R}_n$ for  different ensembles  are expected to undergo similar evolution in  terms of $\Lambda_e$ if subjected to same global constraints. The latter ensures that each ensemble can evolve from statistically same initial condition e.g an ensemble of uncorrelated eigenvalues (corresponding to Poisson statistics).   We note that the set of constants of evolution i.e $t_2, \ldots , t_M$ need not be same for different ensembles; $\Lambda_e$ then describes the evolution of  ${\mathcal R}_n$  for  different ensembles on parallel curves in parametric space.


Following from eq.(\ref{y}) and eqs.(\ref{alme0}, \ref{alme1}),  $\Lambda_e$ is a function of many ensemble parameters and can take a same value for their different combinations i.e for different ensembles.  
As all statistical measures can in principle be derived from ${\mathcal R}_n$, their analogy across different ensembles is expected if the respective  $\Lambda_e$ values are equal and their initial states are statistically analogous. (For examples,  the initial spectral statistics of different ensembles can belong to Poisson universality class if each consists of  sparse complex matrices even if the type  of their sparsity differs.  This is further explained in \cite{psgs} by examples).    The ensemble parameters  for the statistical spectral analogy of two ensembles say  $\rho_1(H)$ and $\rho_2(H)$ can be obtained by  invoking the condition $\Lambda_{e,  \rho_1} = \Lambda_{e,  \rho_2}$, with $\Lambda_{e}$ given by eq.(\ref{alme1}) (assuming their initial state at $\Lambda_e=0$ is same).  

Besides the spectral range,  the dependence of $\Lambda_e$  on the system size (i.e the matrix size $N$ in the ensemble) also plays an important role in the above mentioned statistical dynamics.  As both $(Y-Y_0)$ as well as $R_{1, local}$ can in general  be $N$-dependent (with $N$ as system size),   we have $\Lambda_e \propto N^{\alpha}$ with $\alpha$ arbitrary and  dependent on other system parameters.  A variation of system conditions can cause $\alpha$ to vary, resulting in $\Lambda_e$ varying from $0 \to \infty$.  This in turn results in $R_n$ and other spectral fluctuation measures undergo a crossover from  the initial statistics to Ginibre universality class if $N$ is finite.  In the limit $N \to \infty$,   a change of $\alpha$ may however lead to an abrupt transition from initial  (for $\alpha < 0$) to Ginibre universality class (for $\alpha >0$).  An additional universality class, different from both stationary limits, could however occur if $\alpha=0$ i.e if the underlying system conditions conspire to give rise to an $N$-independent, non-zero and finite $\Lambda_e$. The critical value, referred hereafter as  $\Lambda^*$,  indicates the existence of a non-stationary universality class in infinite size limit and  characterizes the critical spectral statistics different from the two end points i.e initial and Ginibre universality class.   As the  existence of $\Lambda^*$  depends significantly on the appropriate combination of underlying system conditions,  it can differ from system to system and may not even exist for some systems (e.g. it could coalesce with either $\Lambda_e=0$ or $\infty$).   
 
 We note the existence of a critical spectral statistics in a specific non-Hermitian ensemble was discussed in \cite{gnv}; the criticality here was shown to be similar to a specific Hermitian case (i.e  based on the exponential decay of spectral correlations beyond Thouless energy and resulting in a asymptotically  linear behavior  of the number variance with slope less than one). The characterization of the criticality by $\Lambda^*$ mentioned above however is more general and is applicable to any non-Hermitian system described by a multiparametric Gaussian ensemble of complex matrices.  This is further explained in next section by a numerical analysis of three different ensembles of complex matrices. 
 

\section{Numerical analysis of spectral fluctuations}

To achieve the objectives mentioned in section I,  we numerically analyze three ensembles of complex random matrices with independent, Gaussian distributed entries with zero mean but different functional dependence of their variances.  The ensemble density $\rho(H)$  in each case is given by eq.(\ref{pdf}), with $x_{kl;s}=0$ for all $k,l$ and $s$ indices but  $y_{kl;s}$ varying among elements.  The  three ensembles considered here are same as used in \cite{psgs} for numerical analysis of ensemble averaged spectral density; the analysis of their spectral fluctuations here  complements the previous study. The choice also helps to retain a knowledge of the spectral background on which the  fluctuations are imposed (required for unfolding purposes).  

\subsection{Details of the ensembles}

 To avoid a repetition,  here we  mention only the details of the ensembles necessary for present analysis (with more details given in \cite{psgs})

(i) Ensemble of complex matrices with constant diagonal to off-diagonal variance ratio (BE): 
\begin{eqnarray}
y_{kl;s}= {1\over 2} \left({1+{N\over b}}\right)^2,  \hspace{0.1in}  y_{kk;s}={1\over 2}.
\label{be1}
\end{eqnarray}

(ii) Ensemble of complex matrices with power law off-diagonal variance (PE): 
\begin{eqnarray}
y_{kl;s}={1\over 2} \left( 1+ \left({|k-l|\over b}\right)^2 \right)^2,  \hspace{0.1in}  y_{kk;s}={1\over 2} 
\label{pe1}
\end{eqnarray}

(iii) Ensemble of complex matrices with exponential off-diagonal variance  (EE):   
\begin{eqnarray}
y_{kl;s}= {1\over 2} \, {\rm exp}\left({|k-l|^2\over b^2}\right) ,  \hspace{0.1in}  y_{kk;s}={1\over 2}.
\label{ee1}
\end{eqnarray}


We note that the  ensemble density in eq.(\ref{be1}) is same as the Brownian ensemble that appears as a nonstationary perturbed state   when an ensemble  of diagonal matrices is perturbed by a Ginibre ensemble of complex matrices.  Similar Hermitian matrix ensembles,  arising as the nonstationary states  due to perturbation of a diagonal ensemble  by a Gaussian unitary ensemble (GUE)  have also been extensively studied in past;  their ensemble density  is  analogous to the Rosenzweig-Porter ensemble of complex Hermitian matrices too \cite{psrp}.   This motivates us to refer the  case described in eq.(\ref{be1})  as a Brownian ensemble appearing between Poisson and Ginibre limit (or just BE for short).
Further Hermitian ensembles, similar to eq.(\ref{pe1} and eq.(\ref{ee1}) and known as power law random banded ensembles (PRBME) and exponential ensembles respectively,  have also been studied in past  and have been shown to be relevant for many localization to delocalization transitions of many body Hamiltonians \cite{psrp}.    The choice of similar  ensembles in the present work makes a comparative analysis  of the fluctuation measures  in Hermitian and non Hermitian cases feasible, thereby providing some insights for the theoretical formulations for analogous cases,  and is one of the motivations  for their  consideration in the present work.



As indicated above,  each ensemble depends on at least two system parameters, namely, $b$ and system size $N$. Contrary to BE, the off-diagonals  in PE and EE  depend on the distance $r=|k-l|$ from the diagonal too,   with their strength decaying with $r$ as a power-law and exponentially, respectively.  This makes faraway off-diagonals  relatively negligible with respect to diagonals,  making a typical  matrix of the ensemble effectively sparse,  with sparsity  in EE stronger than PE.
 To understand the system dependence of the spectral correlations,  we exactly diagonalize thousands of matrices of each ensemble for many $b$ values and four system sizes $N= 256, 512, 1024,  2048$.

In a previous study \cite{psgs}, we presented a detailed theoretical analysis of the average spectral density $R_1$  for the ensemble (\ref{pdf}) and  derived its formulation in terms of $Y$ (eq.(\ref{y})) for a fixed set of global constraints.  The theoretical claim was verified by a detailed numerical analysis of $R_1$ for BE, PE and EE;  the required $Y$ parameter in each case were obtained by the substitution of their ensemble parameters in eq.(\ref{yt2}), and choosing $b=1/N$ as the initial condition for each ensemble.   
The  latter  along with eq.(\ref{y}) gives
\begin{eqnarray}
Y-Y_0 &=&  -2 \, {\rm ln} \;{ (1 + N/b) \over (1 + N^2)}  \hspace{1.9in}  (BE),    \label{be2}\\
&=& - {2 \over N^2} \sum_ {r=1}^N  \, (N-r) \, \ln\left({1+r^2/ b^2 \over 1+N^2 r^2 } \right) \hspace{0.6in}   \label{pe2} (PE),  \\
&=&      {(b^2 N^2-1) (N^2-1)\over 12 \, b^2} \hspace{1.6in}  (EE). 
 \label{ee2}
\end{eqnarray}
 As indicated by study \cite{psgs},  the increasing repulsion of the eigenvalues with increasing $b$ causes eigenvalues to distribute uniformly on the complex plane with average density $R_1(r, \theta)$ approaching a constant.  While the limit $b \to \infty$ corresponds to  the Ginibre statistics for each ensemble, the limit $b \to 0$ the Poisson statistics.   We note that $(Y-Y_0) \ge 0$ for each case, increasing from $0 \to \infty$ as $b$ varies from $1/N \to N$. We  also note,  that $(Y-Y_0)$ for EE case becomes very large for all  $b$-values above $b = 1/N$.   As  the average spectral density  $R_1(r, \theta)$ is governed by $Y-Y_0$,  this  implied its rapid transition from the initial state to Ginibre limit and was numerically confirmed too in \cite{psgs}. 
 


 As discussed in previous section,  contrary to $R_1$,  ${\mathcal R}_n$ for $n >1$, and, thereby other measures of the  spectral fluctuations over the ensemble,   evolve as a function of  $\Lambda_e$ from an arbitrary initial condition and for a fixed set of evolution constants $t_2, \ldots, t_M$.  Although not required for our fluctuation analysis,  we note, for completeness sake,  that (i) $M=2 N^2-1$ for each of the three ensembles, (ii) eq.(\ref{pe1}) and eq.(\ref{ee1}) can be rewritten as $y_{kl;s} = {1\over 2}\left(1 + {t_r^2 \over b^2} \right)^2$ and $y_{kl;s} = {1\over 2}{\rm exp} \left( {t_r^2 \over b^2} \right)$ with $r \equiv |k-l|$,  respectively.  Inverting eq.(\ref{pe2}) and eq.(\ref{ee2}),   $b$ in preceding relations can be replaced by $Y-Y_0$. This helps to express each $y_{kl;s}$ as a function of $Y$ and $t_r$ only,   and thereby identification of the constants of evolution for PE and EE case as arbitrary functions of $|k-l|$.  The constants for BE case however  can be chosen as all equal e.g.  either $1$ or $N$.   Different constants of evolution imply the evolution of three ensembles on  different  curves but parallel in the parametric space.
 
   
 To numerically verify our theoretical claim,  we need $\Lambda_e$  expressed explicitly in terms of the  system parameters of the three ensembles.  As in \cite{psgs}, here again we choose $b=1/N$ as the initial condition for each ensemble; this ensures, in large $N$-limit, that the initial spectral statistics in each case belongs to the Poisson universality class on complex plane.  The choice also retains $Y-Y_0$ for each ensemble  same as those used in \cite{psgs}  for the average level density analysis (given above by eq.(\ref{be2}, \ref{pe2}, \ref{ee2})).   But a determination of  $\Lambda_e$ from  eq.(\ref{alme1}) also requires a prior information about $R_{1,local}(e)$ at the spectral range $e$ of interest; as mentioned in previous section,  a complete understanding of the latter  is missing so far.  This  handicaps us, in the present work,  from theoretical prediction of $\Lambda_e$, leaving its numerical determination as the only option. As discussed below  our numerical analysis suggests  
 \begin{eqnarray}
 \Lambda_e \propto \left(\log b - c\right)
\label{almen}
\end{eqnarray} 
with $c$ as an ensemble dependent constant.   It is worth emphasizing here  that $b$ appears differently in each ensemble and has different effect on the relative strengths of the matrix elements. Indeed a change of $b$ in BE leaves the off-diagonals  unchanged relative to each other but changes them with respect to the diagonals.  On the contrary, for both PE and EE cases, the elements in any two principal off-diagonals change only relatively as $b$ changes .
 
 



The form in eq.(\ref{almen}) for $\Lambda_e$ can indeed be justified for BE and PE:  from eq.(\ref{be2}) for BE,  we have $Y-Y_0 \approx {\rm ln} \, b - c$ for $b < 1$ where $c ={\rm ln}{N \over (1 + N^2)}$;  this would be consistent with eq.(\ref{alme1}) if $R_{1, local}$ is constant for this case.     A similar behavior for $Y-Y_0$ from eq.(\ref{pe2}) for PE  can be confirmed by the computational intelligence tools (Wolfram Alpha).  
 Surprisingly,   eq.(\ref{almen}) for $\Lambda_e$  seems to work well for EE too (although  not directly obvious from eq.(\ref{ee2})).  Indeed,  as can be seen from eq.(\ref{ee2}),  $Y-Y_0$ is almost independent of $b \sim N^{\alpha}$ with $\alpha \ge 0$.  As showed in \cite{psgs},  the average level density $R_1$ also showed a rapid crossover from $\alpha =-1$ to $\alpha \sim 0$).   But as discussed below,  the local spectral correlations for EE  vary smoothly in terms of the parameter $\left(\log b - c\right)$ even when $\alpha \sim 0$.  Following the definition in eq.(\ref{ee2}),  the  $\Lambda_e$-form in eq.(\ref{almen})  for EE case can then  originate only from $R_{1, local}$.


\subsection{Non-ergodicity and Non-Stationariness of the Fluctuations}

The ergodicity in a random matrix ensemble implies the equivalence of the average  of a measure when its behavior is considered along a single spectrum  with that obtained over an ensemble for a fixed spectral point (similar to equivalence of the averages over time and the phase space of standard statistical physics).   The existence of the ergodicity for  the average spectral density as well as its fluctuations for  a basis-invariant  random matrix ensembles e.g. Gaussian orthogonal ensemble,  Gaussian unitary ensemble etc was first proved in \cite{pan}.  

The stationarity of the spectrum however corresponds to an analogy of the local fluctuations at any two  arbitrary spectral points,  equivalently,  translational invariance in case of the spectrum along a real line and both translational as well as rotational invariance in case of complex plane.  
As the statistical measure of  an ergodic ensemble are translationally/ rotationaly invariant along the spectrum, it is also referred as a stationary ensemble.   Indeed both the ergodicity as well as stationarity  of a random matrix ensemble are signatures of the  underlying  basis: an invariance in the latter  permits the eigenfunctions to be ergodic and spectral measures to invariant along the spectrum axis.


Intuitively the average spectral density, of a non-stationary ensemble  is expected to be non-ergodic: its behavior  on an average over a single spectrum  differs from that obtained   over an ensemble while keeping the spectral point fixed \cite{pan, bg}.    Indeed the theoretical study \cite{psgs}, supported by numerical evidence,  has already indicated that  the average spectral density of the ensemble (\ref{pdf}) is in general non-ergodic. This motivates the natural query: whether the local fluctuations  imposed on a non-ergodic background will retain non-ergodicity even after unfolding?
As discussed in \cite{bg,  pscons2, haak,  brody},  the conditions for ergodicity differ from one measure to another and  its lack for average spectral density need not hinder its existence for the spectral fluctuations.   We note that,  in case of the latter, ergodicity can at best be defined locally: it exists if the local fluctuations {\it in a range $\Delta z$ around an arbitrary spectral point "$z$"  for a single matrix }  are analogous to those {\it over an   ensemble  at the point "$z$"}.   We refer such a spectrum as "locally ergodic" and emphasize  that it differs from the  non-stationarity of the fluctuations.


Besides improving statistical accuracy and thereby enabling a meaningful comparison  of the local fluctuations, the information about the local ergodicity  is needed,  in the present context, for another reason too.   In case of a stationary ensemble,  the local fluctuations of the spectral density depend only on its spectral average, equivalently,  its ensemble average due to ergodicity; the unfolding maps the latter to a constant and thereby mapping  the local fluctuations on a constant background.
But as indicated by eq.(\ref{rnp2}), the local fluctuations in a non-stationary ensemble depend on both $R_{1,local}(z)$ as well as $Y-Y_0$ (from the definition  of $\Lambda_e$) and while the unfolding removes their explicit dependence  on the location $z$ on the complex plane,  the latter is still retained, implicitly,  through $\Lambda_e$.  Indeed this is the  ensemble's non-stationarity  manifesting through non-stationarity of the local fluctuations even after unfolding. 
 As a consequence,  a spectral averaging e.g. of the spacings over a finite spectral width, say $z-\delta z, z+\delta z$, can  lead to mixing of statistics with different $\Lambda_e$ unless $\delta(z)$ is very small or the local average density is almost constant (thereby leaving $\Lambda_e$ almost constant).  While this can be avoided by considering just the ensemble averaging i.e by taking a single (or two) unfolded spacing $S_k$ at spectral point $z$ from a  matrix  and averaging over all the matrices in the ensemble, this in turn requires a large ensemble size for accuracy purposes and therefore huge computational time.  
 
 The existence of  local  ergodicity in the spectrum helps in the above context:  the unfolded spacing averages over an ensemble at a spectral point, say "z" can be approximated by the double averages i.e those over an ensemble as well as in an optimized spectral range  around "z" (small enough to keep $\Lambda_e$ almost constant).  For example,  in the spectral regions  where $R_1(z)$  varies very slowly,  it is  possible to choose an optimized range  in the neighbourhood of $z$, sufficiently large for good statistics but keeping  a mixing of different statistics at 
minimum.   (Note this is usually not the case for the regions e.g.  edge with sharp change of $\Delta(z)= (R_{1}(z))^{-1}$; the latter leads to a rapidly changing  $\Lambda_e$  and it is numerically difficult to consider a spectral range with an appropriate number of levels  without mixing of different statistics).  


 As shown in figures 1-3 of \cite{psgs},   the eigenvalues for each of the three ensembles  are distributed over a complex plane with a non-ergodic, $Y$-dependent density and, as a consequence,  their rescaled (unfolded) spacings can vary (i) from one spectral range to another at a fixed $\Lambda_e$, or (ii) with $\Lambda_e$ at a fixed spectral range.  (The corresponding behaviour for Hermitian ensembles in discussed in detail e.g.  in \cite{ptche}).  The nearest neighbour spacing distribution (NNSD) $P_e (S, \Lambda_e, e)$ for the rescaled eigenvalues  therefore depends on the variables $S, \Lambda_e$ as well as $e$, with $S$ as a nearest neighbour spacing in the neighbourhood of a point $e$ on the spectral plane with 
 $\Lambda_e$ as the ensemble complexity parameter (definition of $P_e$ given in {\it appendix B}).   Here the unfolding of the spacing is achieved by rescaling  it  by the local mean spectral density numerically defined as \cite{hkku, prosen,  swhjy}
\begin{eqnarray}
R_{1,unf}={n \over \pi d^2_{n, k}}
\label{unf}
\end{eqnarray}
 where $n$ is an optimized number of the eigenvalues, sufficiently larger than unity but small compared to the matrix size $N$ and $d_{n, k} \equiv |z_{n+k} -z_k|$ is the $n^{th}$ nearest neighbour distance from $z_k$ (before unfolding). The rescaled spacing are then defined as $S_k = d_{1,k} \sqrt{R_{1,unf}}$.  (We note that the function in eq.(\ref{unf}) corresponds to  an  unfolding of each spectrum locally and is used in many previous studies \cite{hkku, prosen,  swhjy}.   Alternatively, it can also be done by the approximate  $R_1(r, \theta)$ formulation derived  in our previous work \cite{psgs} but the final results remain affected).
 
 



 
 To seek presence/ absence of the local spectral  ergodicity,  we consider two different averaging approaches  for each of the three ensembles.  
 The first averaging route, referred as the {\it ensemble averaging},  is as follows: We recall that, each eigenvalue on the complex plane  has more than two neighbors; (this is contrary to the Hermitian case in which the eigenvalues are confined on real lines and each eigenvalue has two nearest neighbors). 
 We first choose one of the eigenvalues  at $e \sim 0$ from a matrix,  choose its spacing with nearest neighbor.  As the exact eigenvalue at $e \sim 0$  or its nearest neighbor spacing indeed need not be same for different matrices,   we study the distribution of such spacings over an ensemble of 2500 matrices; thus the total number of eigenvalues used for the analysis are $2500$.  (This corresponds to using the  definition  $P(S,  \Lambda_e) = \langle \sum_k \delta(S-S_k) \rangle$ with $\langle \rangle$ as the ensemble average).
 The second route, based on both spectral as well as ensemble averaging and referred here as {\it spectral-ensemble averaging}  is as follows:  we consider $10\%$ spacings around $e=0$ for each matrix in the ensemble, obtain their distribution for the matrix and then average it over the ensemble; (for the matrix size  $N=1024$, this corresponds to $ \sim 100$ spacing taken from a single matrix with ensemble size $M =25$).   (This corresponds to using the  definition  $P(S,  \Lambda_e) = \langle \; \overline {\sum_k \delta(S-S_k)} \; \rangle$ with $\overline x$ as the spectral average).
 The number of levels used in each case are kept same,  to avoid non-physical deviations arising from  e.g. finite size sample errors.  
 
 Figure 1 displays a comparison of the $P(S,  \Lambda_e)$,  obtained by the two averaging approaches,  in the neighborhood of the region  $e \sim 0$  for three $b$-values (equivalently for three $\Lambda_e$) for each of the three ensembles.  
 While  a slight deviation between the two averagings is clearly visible from figure 1 for each of the three ensembles for $b \ll N$,    it survives, albeit rather  weakly,  even for the Ginibre limit ($b=N$ case) of each ensemble  too.   Theoretically however the Ginibre ensemble as well as its local fluctuations are  expected to be stationary.   This suggests  the deviation  to be an artefact of the finite sample errors.  Indeed the two types of averaging are  almost in  agreement for each  ensemble,  thereby suggesting the local ergodicity of the fluctuations.  But the latter does not imply stationarity of the spectrum.  This can further be illustrated by a consideration of the Ginibre ensemble directly; as the latter's spectrum is stationary,  the distribution of $P(S,  \infty)$ is expected to be independent of the location $"e"$ and is indeed confirmed in  figure 2.  In contrast, as indicated by our numerics,  the deviation for BE, PE and EE are sensitive to the spectral location of the single spacing thus indicating non-stationarity of the spectrum (corresponding figure not included here).   Further the analogy of the two averages for the Ginibre ensemble for $1^{st}$ spacing is visibly better than PE and EE cases for $b=N$; this suggests the fluctuations reaching an almost Ginibre limit,  but not exactly, for them. 
 

\subsection{Lack of Sensitivity to Multiple System Parameters: emergence of Complexity Parameter}

As discussed in section II,  the fluctuations are governed by a single functional $\Lambda_e$ describing a collective influence of the system conditions and not their individual details.   $\Lambda_e$ can however be changed by changing $Y$ i.e by changing the system conditions.   Thus  the unfolded correlations  can change even at a given spectral point   "$e$"  if the system parameters and thereby ensemble parameters are varied.   Based on our theoretical prediction,   a variation of the ensemble parameters   should lead to an analogous crossover in a spectral fluctuation measure for different ensembles if they share statistically same initial  and final endpoints.   To verify the above,  here we statistically analyze each ensemble by varying the parameter $b$ and thereby $Y-Y_0$.   As the latter changes the average mean level density $R_1$ too,   
this affects $\Lambda_e$ both explicitly as well as through $R_1$. 

Contrary to a stationary spectrum, the local correlations in our case are not translational or rotational invariant i.e vary on the complex  spectral plane from one point to another for a  fixed set of system conditions i.e for a fixed $Y$.     An unfolding of the levels maps them, locally, to a constant background.  As this renders  the fluctuations locally ergodic, one can consider levels within an optimized spectral range without mixing the statistics.  But as the fluctuations of the unfolded eigenvalues are nonstationary  (governed by $\Lambda_e$ that varies along the spectrum),  it is necessary to choose an optimize range $\Delta e$  that contains sufficient number of eigenvalues to give good statistics but all with almost same $\Lambda_e$.   
We analyze  10${\% }$ of the total eigenvalues taken from a range $\Delta e$ centred at $e \sim 0$ for $N=1024$ and for an ensemble size of 2500 matrices; this gives total $\sim 10^5$ eigenvalues for the analysis for  each ensemble which are unfolded by $R_{1, unf}$ given by eq.(\ref{unf}).


As the non-stationarity of the spectrum  renders a definition of the long range spectral measure not very clear (unfolding not leaving them locally stationary),  here we consider two fluctuation measures, both indicators of short range spectral behavior. 



{\it Spacing Distribution:}  While no theoretical formulation is currently available for the $P_e(S,  \Lambda_e)$ intermediate between Poisson and Ginibre limit,  its approximation by $R_2$  ({\it appendix B}) suggest,  along with eq.(\ref{rna3}),  a  smooth crossover governed by $\Lambda_e$.   Thus the form of  $P_e(S,\Lambda_e)$  is expected to be intermediate between  Poisson ($\Lambda_e=0$, eq.(\ref{pspoi})) and Ginibre universality class ($\Lambda_e=\infty$,  eq.(\ref{psgin}))  if  $\Lambda_e$ is  finite and non-zero.   
Figure 3 displays the $P_e(S, \Lambda_e)$  behavior for each of the three  ensembles for many $b$-values, equivalently many $\Lambda_e$ values obtained from eq.(\ref{almen}).   
Here again,  the spacings considered for the analysis consist of only $10 \%$ eigenvalues from the neighborhood of $e \sim 0$ for a fixed matrix size $N=1024$ and for an ensemble of  $2500$ complex  matrices.  As clear from the figure,  the statistics undergoes a crossover from Poisson to Ginibre statistics as $\Lambda_e$ varies.    We recall that the off-diagonal variances in PE and EE depend on additional system conditions; this manifests through the difference of their decay types, mentioned in the beginning of section III.A) and thereby $Y-Y_0$.
Although the latter has different mathematical form for the three ensembles,  and,  its variation effectively implies variation of more than one system parameter for PE and EE (as clear from  eq.(\ref{be2},  \ref{pe2}, \ref{ee2})),  $P_e(S, \Lambda_e)$ seems to evolve between two stationary limits in an analogous way for the three ensemble.  (A slight deviation of the peak towards left  for BE cases seems to be an artefact of unfolding issues with local mean level density changing more rapidly than PE and EE).  

%

The numerical results shown in figure 3 encourage us to conjecture a form of $P_e$, based on a single parametric interpolation between two stationary limits i.e  Poisson at $\Lambda_e =0$ and Ginibre at $\Lambda_e =\infty$, 
\begin{eqnarray}
P(S, \Lambda_e) = A \; S^{B}  \; {\rm e}^{-C \; S^2}
\label{psfit}
\end{eqnarray}
subjected to  the normalization condition $\int_0^{\infty} P(S) \; {\rm d}S =1$ and unit mean level spacing condition  $\int_0^{\infty} S \; P(S) {\rm d}S =1$.  Here $B=1$ corresponds to Poisson limit (on a plane) and $B=3$ gives the approximate Ginibre limit \cite{haak,  hkku}.   
To justify the single parameter formulation,   it is desirable to express the fitted parameters $A, B, C$ as  as functions of $\Lambda_e$.    As given in table 1,  these parameters can indeed be recast as functions of $\Lambda_e$. 

We note, 
that previous numerical studies of $P(S; \infty)$ for non-Hermitian case (i.e Ginibre ensemble) indicate an important difference from the Hermitian one (i.e Gaussian unitary  ensemble): 
although, in the latter case,   the Wigner surmise for $P(S)$ is exact for $N=2$ case only,  it has been shown to be well-applicable for large $N$ case too.    The large $N$ limit of $P(S)$ for Ginibre ensemble however deviates significantly from the $N=2$ result \cite{hkku}; (see eq.(\ref{psgin}) of {\it appendix} B); this suggests emergence of new spectral correlations on the plane as $N$ increases.  As displayed in figure 4,   the good agreement of our  numerical results with eq.(\ref{psfit})  indeed supports the above conjecture.  The numerically  fitted values for the parameters $A, B, C$   and their suggested $\Lambda_e$ dependence for each ensemble is given in table 1.   Here again the definition $\Lambda_e=\log b-C$ as the crossover parameter seemingly apply quite well.



\begin{table}
\caption{\bf The table gives the  fitted parameters $A, B, C$ for eq.(\ref{psfit} used in figure 4 for BE, PE and EE respectively.   Based on the numerics,  the fitted parameters  can be recast in the suggestive forms as a function of $\Lambda_e$; this is given in the second row for each case 
The second row for each case gives the suggested $\Lambda_e$ dependence of $A, B, C$. }
\begin{tabular}{|c|c|c|c|c|}
\hline
\hline

{\bf Case}      & $\Lambda_e$    &      A    & B         &  C       \\
BE  & $\log b + 3.79$  & $6.78 (\log b + 3.15)$,   &  $2.1 ( \log b + 3.79)-0.31$,   & $ 0.92 \log b + 3.79$;  \\
&  & $ \sim 6.78 \Lambda_e$,  &  $ \sim 2.1 \Lambda_e,$  & $\sim \Lambda_e $ \\
\hline
PE  & $\log b + 1.48$  & $7.39 (\log b + 1.03)$,   &  $3.27 (\log b + 1.2)$,  & $1.43 (\log b + 1.48)$\\
&    & $ \sim 7.39 \Lambda_e$,  & $ \sim 3.27 \Lambda_e$,  & $\sim  1.43 \Lambda_e $ \\
\hline
EE &$\log b -1.17$ & $6.01 (\log b -1.88)$,   &  $2.06 (\log b -1.17) -0.46$,  & $ \log b-1.17$ , \\
      &  &  $\sim 6.01 \Lambda_e,$  &  $\sim 2.06 \Lambda_e,$, & $   \Lambda_e$\\

\hline
\hline
\end{tabular}
\end{table}


\vspace{0.1in}

{\it Spacing Ratio Distribution:} 


As briefly discussed in {\it appendix C},  the unfolding related technical  issues and errors can be avoided by the consideration of  a dimensionless short range fluctuation measure.  i.e the nearest neighbour spacing ratio,   previously introduced for the real eigenvalues of Hermitian ensemble \cite{huse, abgr}.
With eigenvalues of a complex matrices distributed on a complex plane,  a generalization of the  spacing ratio measure,  referred as the distribution of the complex spacing ratios,   was later introduced in \cite{prosen}  for non-Hermitian ensembles.   

A complex spacing ratio $Z_k$ is defined as the ratio of the nearest neighbour spacing  and next nearest neighbour spacing of a given eigenvalue $z_k$ on the complex plane 
\begin{eqnarray}
Z_k = {z_k^{NN} - z_k \over z_k^{NNN}-z_k}. 
\label{zk}
\end{eqnarray}
with   $0 \le Z_k \le 1$, $z_k^{NN} $ and $z_k^{NNN}$ as the nearest neighbor and next nearest neighbor eigenvalue of $z_k$. The probability density  $P_z(Z)$ of finding a spacing ratio $Z$   over an ensemble for a specific spectral range is then free of unfolding issues and  contains information not only about the radial correlations but  the angular ones too \cite{prosen}.   The latter can also be seen directly from the marginal distributions of the  radial and angular parts of the spacing ratio; writing $Z= r \; {\rm e}^{i \theta}$,   we have 
 \begin{eqnarray}
 \rho_r(r) &\equiv & \langle P_z(Z) \rangle_{\theta}  = \int {\rm d}\theta \; r \; P_z(Z), \\
\rho_{\theta} (\theta) & \equiv & \langle P_z(Z) \rangle_{r} = \int {\rm d}r \; r \; P_z(Z).
 \label{rhort}
\end{eqnarray}  

For uncorrelated eigenvalues distributed on a complex plane,  the spacing ratios are isotropic and consequently \cite{prosen} 
\begin{eqnarray}
\rho_{r, poi}(r) \propto \; r \; \Theta(1-r)
\label{poi1}
\end{eqnarray}
with subscript "poi" as the reference to Poisson statistics.  No closed form expression for lage $N$ is however  known for Ginibre ensemble of complex matrices (referred as GinUE in \cite{prosen}). 
Although an exact result is obtained for $N=3$ in \cite{prosen} but, contrary to Hermitian case,   it is not a good approximation in large $N$ limit; (we recall that  exact for $N=2$ for Hermitian matrices  well approximates large $N$ behavior too).
As discussed in \cite{prosen}, the leading order expansion in powers of $r$ in large $N$-limit 
yields  $\rho_{r, gin}(r) \propto \; r^3$  (with subscript "gin" as the reference to Ginibre statistics)
valid only for spectral range around $r=0$.   In the regime intermediate to Poisson and Ginibre ensemble,  however,  no  theoretical results are currently available for either $P_z(Z), \rho_r(r)$ or $\rho_{\theta}(\theta)$.

Figure 5 displays the behavior for $\rho_r(r)$ and $\rho_{\theta}(\theta)$ for the three ensembles in the bulk of the spectrum near $e \sim 0$ for many $b$-values. To numerically calculate $\rho_r(r)$, we again consider $10\%$ spacing-ratios around $e=0$ for each matrix in the ensemble, obtain the distribution of its radial part for the matrix and then average it over the ensemble; (for the matrix size  $N=1024$, this corresponds to $ \sim 100$ spacing taken from a single matrix with ensemble size $2500$). Similarly $\rho_{\theta}(\theta)$
is obtained by considering the distribution of the angular part for the spacing ratios for each matrix and subsequently averaging it over the ensemble.  Besides reconfirming the information contained in figure 3,  i.e a smooth crossover from Poisson to Ginibre with increasing  $b$ and thereby $\Lambda_e$ for each ensemble,   the left panel of the figure 5  also indicates 

(i) an almost power law variation of $\rho_r(r)$ with $r$ for a fixed $\Lambda_e$,  

(ii) an increasing $b$ resulting in decrease of $\rho_r(r)$  for $r < 0.5$ but its increases  for $r > 0.5$,  

(iii)  while the curves for intermediate $b$ values lie between Poisson and Ginibre limit in both regions $r < 0.5$ and $r > 0.5$, they seem to converge  in the neighborhood of $r \sim 0.5$.  More clearly,  as $\rho(r)$ at $r \approx 0.5$ is almost same  irrespective of $b$,  equivalently $\Lambda_e$,  this indicates $\Lambda_e$-insensitivity of $\rho(r)$ at $r \approx 0.5$ and suggest, in limit $N \to \infty$, an existence of the critical point of statistics at $r \sim 0.5$. We note, from eq.(\ref{zk}),  this corresponds to a specific relation between two nearest neighbors of a given eigenvalue. 

As displayed in the right panel of figure 5,  $\rho(\theta)$ drops rapidly  near $\theta=0$ as $b$ and thereby $\Lambda_e$ increases; clearly the drop occurs due to increasing level repulsion with $\Lambda_e$.  But,  in comparison to $\rho(r)$,  the variation in $\rho(\theta)$ with $\Lambda_e$ is more conspicuous; this indicates the important role played by angular correlations in the statistics.  We also note that the case $\Lambda_e=0$ is indeed almost constant and is therefore  in agreement with theoretical prediction for Poisson universality class (on the complex plane).

\subsection{Universality and Criticality in Spectral Statistics}

 Based on eq.(\ref{rnp1}),  our theoretical claim is the following: different ensembles subjected to same global constraints  are expected not only to undergo similar evolution of the  spectral statistics but also to display same statistics if their $\Lambda_e$ values are same.  It will be instructive to verify the claim for BE, PE and EE where, due to different variance structures, it is not at all obvious as to why their statistics be analogous for some specific set of system parameters.    
The $\Lambda_e$ values for BE, PE and EE, where their statistics are predicted to be analogous,   can in principle be determined by invoking the condition $\Lambda_{e, BE} = \Lambda_{e, PE}= \Lambda_{e, EE}$, with $\Lambda_e$  given by eq.(\ref{alme1}).  Due to a  current lack of understanding about  $R_{1,local}(e)$,  however,  here we consider a more general route: 
instead of comparison at a single $\Lambda_e$ value,  we compare  the local fluctuations for the three ensembles for full crossover,  from $b =1/N \to b=N$,  as a function of $\Lambda_e$ defined in eq.(\ref{almen}); an agreement would  indicate that the analogy is not just an accidental coincidence for some $b$ value and  indeed exists with $\Lambda_e$ defined by eq.(\ref{almen}). 

The parametric evolution of the spectral fluctuations between two stationary limits can be best studied by analyzing  an average measure.  A traditionally used measure in this context   is  the cumulative nearest neighbour distribution $I(x, e, \Lambda_e) =\int_0^{x} P_e(S, e; \Lambda_e) \; {\rm d}S$ with $x$ as an arbitrary point.  Alternatively,  a related measure, particularly useful for the phase transition studies, is the relative behavior of the tail of the nearest-neighbour spacing distribution $P_e$, defined as 

\begin{eqnarray}
\gamma (x; e,  \Lambda_e) = \frac{I(x, e; \Lambda_e)-I(x, e; \infty)}{I(x, e; 0)- I(x,  e; \infty)}
\label{alp}
\end{eqnarray}
We choose $x$ as one of the two crossing points, referred as $x_1, x_2$, of $P_e(S, e; \infty)=P_{gin}(S)$ and $P_e(S, e; 0)= P_{poi}(S)$ (here the subscripts $gin$ and $poi$ refer to the Ginibre and Poisson cases respectively): $x_1 =0.656978$ and $x_2=1.446259$.    As obvious, $\gamma=0$ and $1$ for the Ginibre and Poisson limit respectively  and a fractional value of $\gamma$ indicates the probability of small-spacings,  and thereby short range correlations,  different from the two limits.  In limit $N \to \infty$, a $\gamma$ value different from the two end points is an indicator of a new universality class of the spectral statistics and therefore its critical point.   


For comparison, we consider the $\gamma$ behavior with respect to a parameter,  namely, $\tau \equiv \log(b)-C$ with $C=0, 1.8, 4.6$ for BE, PE and EE respectively;  we note that, from  eq.(\ref{almen}),  $\tau$ is directly related to $\Lambda_e$. 
 Figure 6  shows a comparison  of the $\tau$-variation of $\gamma_1 \equiv \gamma(x_1, e; \tau)$ and $\gamma_2 \equiv \gamma(x_2, e; \tau)$,  in the neighborhood of  the spectral point $e \sim 0$,  for two $x$-values and  three ensembles for a fixed $N=1024$.
The collapse of  $\gamma(x; \tau )$ behavior for each of the three ensemble onto a single curve confirms our theoretical claim regarding a single parameter based universality of the local fluctuations.    Figure 6 also shows a comparison of the variation of (i)  radial part $\langle r \rangle $,    and (ii)  angular part $-\langle \cos(\theta) \rangle$ of the mean level spacing ratio (with $r_k,  \theta_k$  referring to radial and angle part of the average spacing ratio $Z_k$) for three ensembles.  Here the averages  $\langle r \rangle $ as well as $-\langle \cos(\theta) \rangle$  are numerically obtained by the average of $r_k$ and $- \cos \theta_k$ over an optimized energy range ($10 \%$ eigenvalues around $e \sim 0$) as well as the ensemble of $2500$ matrices.    With $N$ large, this permits us to consider a large number of eigenvalues for the level statistics while ensuring that they belong to same $\Lambda_e$.  The good agreement for the three ensembles,  of the two measures  for entire crossover, once again confirms our theoretical claim regarding the single parameter dependence of the fluctuations.  We note that while $\gamma_1, \gamma_2$ are cumulative spacing distributions, 
 $\langle r \rangle $ and $-\langle \cos(\theta) \rangle$ are  related to a dimensionless measure i.e the  spacing ratio.   Further as the collapse of each fluctuation measure for the three ensembles  occurs when plotted with respect to $\tau = \log(b) - C$; this again strongly suggests $\Lambda_e \propto \tau$.


The illustrations in figure 6 suggest  another important aspect of the fluctuations, namely,  the existence of a critical statistics:  as $\log (b)$ increases,  $\gamma_1, \gamma_2,  \langle r \rangle$  and $-\langle \cos\theta \rangle$ undergo a rapid transition from Ginibre ($\gamma_{1,2}=1$) to Poisson limit ($\gamma_{1,2}=0$); this in turn suggests the existence of a critical statistics near $\log(b) -c \approx -2.5$.  Indeed it is worth noting  while the three ensembles have different variance structures of the matrix elements, the point of inflection seems to be same in terms of the parameter $\log b -C$.   


As mentioned in previous section,  the spectral statistics of a non-stationary ensemble approaches,  in limit $N \to \infty$,   one of three universality classes i.e either to two end points or the critical statistics.   The latter is,  by definition,  size-independent and is  characterized by a size-independent $\Lambda_e$.   We note that a similar criticality of the Brownian ensemble has already been reported for Hermitian cases \cite{psrp}.    
To confirm the existence of a critical point,  it is therefore necessary to analyze the size-dependence of the spectral measures at the critical $b$-values (those resulting in size-independent $\Lambda_e$).   We again numerically analyze  $P(S, e; \Lambda_e)$ for the BE case,  with  a $ b$-value arbitrarily chosen intermediate between Poisson and Ginibre limits,  and,  
for $10 \%$ eigenvalues lying in the neighborhood of $e \sim 0$ and for the matrix sizes $N= 256,  512, 1024, 2048$ with  ensemble size as $10000,  5000, 2500,  1250$ respectively; this gives the  total number of eigenvalues  used for the analysis as $\sim 2.5 \times 10^5$ for each case.   As displayed in figure 7,    $P(S,  e; \Lambda_e)$  behavior  seems to be almost insensitive to the system size $N$. 
This motivates us to consider the measures $\gamma_1, \gamma_2, \langle r \rangle, -\langle \cos \theta \rangle$ for BE  for many $b$-and $N$-values (again using the same spectral sample as mentioned above).   As displayed in figure 6,  the measures turn out to be size-independent for almost entire $b$-range,  thus indicating the statistics as critical for arbitrary $b$ value,  and thereby implying ensemble in eq.(\ref{be1}) itself to be critical.  (The existence of similar infinite family of critical ensembles for Hermitian matrices has already been discussed e.g.  in \cite{psand, psbe, psrp}). 
As the figure indicates,  the $b=1/N$ and $b=N$ cases correspond to   $\langle r \rangle=0.67$ and  $\langle r \rangle=0.75$ respectively. As indicated by $P(S)$-study, all three ensembles approach Poisson limit for $b=1/N$ and Ginibre limit for $b=N$. We note  that the measures 
$\langle r \rangle$ and $-\langle \cos \theta \rangle$ were also used in \cite{swhjy} for phase transition studies in a two coupled non-Hermitian Hatano-Nelson chains of interacting fermions in the presence of a random potential; the study indicated $\langle r \rangle=0.5$ as the Poisson limit.

Figures 8 and 9 display the corresponding behavior of the above measures  for PE and EE case too.  As in the BE case,  here again the collapse of the curves for almost entire $b$-range and  different $N$ values,   for each of the measure, namely, $P(S, \Lambda_e), \gamma_1, \gamma_2, \langle r \rangle, \langle \cos \theta \rangle$,  indicates the criticality  of PE and EE  with their distribution parameters given by eq.(\ref{pe1}) and eq.(\ref{ee1}); (the  Hermitian ensembles with 
similar variance structures are also known to form an infinity family of critical ensembles \cite{psand,  psbe, psrp}). 
The results in figures 7-9 also reconfirm  $\log b$ as the parameter governing the spectral statistics; this in turn strongly suggests $\Lambda_e$ as a function of $\log b-C$.  A comparison with non-zero $C$ values in figure 6 thus suggest $C$ as a ensemble specific constant,  independent of size $N$.

An important aspect of complexity parameter formulation for both Hermitian as well non-Hermitian ensembles  is its role as  the criteria to determine the critical spectral statistics.  The present analysis confirms the validity of the criteria for complex matrix ensembles with   critical spectral statistics appearing between Poisson  and Ginibre universality classes: it occurs  at  the system conditions  that keep  rescaled complexity parameter $\Lambda$  size-independent.  As this occurs for a specific combinations of system parameters,  it existence is not always guaranteed for all ensemble types. 

The existence of critical spectral statistics for a specific variance structure of the matrix elements  irrespective of the global constraints e.g. presence or absence of the Hermiticity indicates a hidden critical structure among complex systems:  clearly a Hermitian ensemble 
perturbed by a non-Hermitian one with a critical variance structure of the  matrix elements  (i.e of type given by eqs.(\ref{be1}, \ref{pe1}, \ref{ee1}) is expected to be critical too in its spectral statistics.  As the latter is in turn known to be related to multifractality of the eigenfunctions (for Hermitian case),  this reflects a special kind of quantum correlations hidden underneath the quantum dynamics of these systems. 



\section{Conclusion}

We numerically analyze various physical aspects e.g.  ergodicity vs non-ergodicity, universality and criticality  of the local spectral fluctuations for non-stationary complex random matrix ensembles representing non-Hermitian many body/ disordered  Hamiltonians with changing system conditions and away from equilibrium.  Based on analysis of three prototypical, basis-dependent ensembles,  subjected to same global constraints but with  their variance structure representing three different types of quantum correlations in underlying basis space,  our results clearly reveal non-ergodic, non-stationary as well as the critical aspects of the spectral correlations. This information is  relevant not only for fundamental interests in such ensembles but also in context of quantum information, neural networks as well as the non-Hermitian many body localization e.g. whether or not a non-Hermitian system can achieve thermalization?  

The most important insight provided by our numerical analysis is following: notwithstanding different variance structures,  we find that the spectral statistics for each ensemble undergoes an analogous crossover from Poisson to Ginibre universality class  in terms of a single parameter $\Lambda_e$ and also  numerically identify the latter.  This indicates a universality in bulk spectral statistics underlying in non-equilibrium regime of non-Hermitian systems and  is consistent with theoretical claims based on complexity parametric approach reported in \cite{psgs, psnh}.  A lack of theoretical formulations of measures  handicaps us from  a comparison of our numerical results with theory but indirect verification is given by analogy of evolving statistics, between two stationary limits, for each ensemble in terms of  $\Lambda_e$ .

For a deeper insight in the  complexity parameter based formulation of the  complex systems, it is relevant to compare present results with those already known in the domain of conservative systems.  In case of the latter undergoing localized to delocalized transition with changing system conditions e.g. disorder or many body interactions,  the Hermitian ensembles representing them  undergo a crossover/abrupt transition of spectral statistics from Poisson to one of the ten stationary universality classes.  As confirmed by previous studies, the transition can be well-described by $\Lambda_e$, the spectral complexity parameter  \cite{psall, psrp}. 
Similar studies on Hermitian ensembles with additional constraints e.g positive definiteness or column constraint have also indicated the existence of non-equilibrium universality classes of spectral statistics, characterized by $\Lambda_e$ \cite{psall1, ptche}. While the basic idea remains same as in Hermitian case,  the nature of the universality classes   in non-Hermitian cases is expected to be different.  While we do have a clear understanding of the ensemble complexity parameter $Y$ for non-Hermitian cases,  a  theoretical formulation of $\Lambda_e$ is still missing.  As in the Hermitian case,  its technical definition is again expected to be dependent on eigenfunction correlation and therefore requires a detailed analysis of the latter.

Besides questions of eigenfunction fluctuations,  our study gives rise to many other queries e.g. can the complexity parameter formulation be extended to non-Gaussian ensembles as well as to structured ensembles with correlated matrix elements? The information is  relevant in order to model more generalized class of  non-Hermitian systems e.g. neural networks where existence of additional matrix constraints can lead to many types of matrix elements correlations. For application to many body non-Hermitian systems it is relevant to analyze as to how the behavior of spectral correlations in the edge differ from the bulk? Another very important question, especially in context of topological mateials is the existence of exceptional points and their effect on the  statistical behavior of the left and right eigenfunctions. The information is needed to understand e.g. the non-Hermitian skin effect \cite{r2} and the absence of conventional bulk-boundary correspondence in topological and localization transitions in non-Hermitian systems  and also for applications to biological neural networks. We intend to answer some of the above queries in near future.


\appendix

\section{Complexity paramter formulation of spectral correlations}

As discussed in \cite{psgs,  psnh},  a change in system conditions can affect the matrix elements $H_{kl}$ as well as the ensemble parameters in eq.(\ref{pdf}),  resulting in mutiparametric evolution of $\tilde{\rho}(H)$ in matrix space.  The transformation of parameters,  with help of its Gaussian form further  leads to single parametric formulation of the evolution equation for $\tilde{\rho}(H)$ with a constant diffusion and  finite drift \cite{psgs, psnh},

\begin{eqnarray}
 {\partial \rho_1\over\partial Y} &=&
\sum_{k,l;s}
{\partial \over \partial H_{kl;s}}\left[
 {\partial \over \partial H_{lk;s}}  +\gamma \; H_{kl;s}  \right] \rho_1 
 \label{rhot1}
\end{eqnarray}
where  $Y$ is given by eq.(\ref{y}),  $\rho_1 = C_2 \; \rho = {C_2 \over C} \tilde{\rho}$  (with $C$ as a normalization constant such that $\int \tilde \rho \; {\rm D}H =1$ and $C_2$ as a function of $y_{kl;s},  \, x_{kl;s}$,  discussed in  {\it appendix A} of \cite{psgs}).

The above equation along with  eq.(\ref{pz}) in turn leads to  a $Y$ governed evolution of $\tilde P_e$  (details discussed in {\it appendix B} of \cite{psgs} and also in   \cite{psnh}), 
 \begin{eqnarray}
{\partial P_1\over\partial Y} &=&\sum_{s=1}^2 \sum_{n=1}^N{\partial \over \partial z_{ns}}\left[{\partial \over \partial z_{ns}}  - 2 \, {\partial {\rm ln} |\Delta_N (z)| \over \partial z_{ns}}+ \gamma z_{ns} \right] P_1
\label{pcmp}
\end{eqnarray}
with $P_1(z_1, \ldots, z_N)$ related to the normalized distribution by $\tilde P_e = C_2 P_1/C$,  with $\Delta_N (z) \equiv \prod_{k < l}^N  (z_k-z_l)$.



An integration of eq.(\ref{pcmp}) over the variables $z_{n+1}, \ldots, z_N$ (as well as their conjugates), multiplying both sides by ${N!\over (N-n) !}$ and subsequently using eq.(\ref{rn})  leads to 
four types of terms given as follows. 
\begin{eqnarray}
{N!\over (N-n) !} \int  \,  {\partial P_1\over \partial Y} \, {\rm  D}\Omega_{n+1} &=& {\partial R_n\over \partial Y}, 
\label{rp0}
\end{eqnarray}
and 
\begin{eqnarray}
{N!\over (N-n) !} \, \int  \,  {\partial^2 P_1\over \partial z_{jr}^2} \,{\rm  D}\Omega_{n+1} &=& {\partial^2 R_n\over \partial z_{jr}^2}  \qquad j \le n, \nonumber \\
&=&  0 \qquad j > n,
\label{rp1}
\end{eqnarray}
with ${\rm  D}\Omega_{n+1} \equiv  {\rm  d}^2 z_{n+1}, \ldots, {\rm  d}^2 z_N$ with ${\rm  d}^2 z \equiv {\rm  d} z \, {\rm  d}^* z$. 

Similarly
\begin{eqnarray}
& &{N!\over (N-n) !} \, \int  \,   {\partial \over \partial z_{jr}} \left( \ln|\Delta_N(z)| \, P_1\right)  \,{\rm  D}\Omega_{n+1}
 = 0  \qquad j > n, \nonumber \\
&=& {\partial \over \partial z_{jr}} \left( \ln|\Delta_n(z)| \, R_n\right) 
\qquad i, j \le n, \nonumber \\
&=& {1\over N-n} \, \sum_{j, r} {\partial \over \partial z_{jr}} \int {z_{jr}-z_{{n+1}r}|\over |z_j-z_{n+1}|} \, R_{n+1}(z)\, {\rm d^2}z_{n+1} \qquad i >n, j \le n, 
\label{rp2}
\end{eqnarray} 
with $R_{n+1}(z) \equiv R_{n+1}(z_1, \ldots, z_{n+1})$.
 and 
 \begin{eqnarray}
{N!\over (N-n) !} \, \int  \,   {\partial \over \partial z_{jr}} \left(z_{jr} \, P_1\right)  \,{\rm  D}\Omega_{n+1}
 &=& {\partial \over \partial z_{jr}} \left(z_{jr} \, R_n\right), 
\qquad j \le n, \nonumber \\
&=& 0  \qquad j > n.
\label{rp3}
\end{eqnarray} 
Using the above relations then leads to eq.(\ref{rnp1}).

\section{ Nearest Neighbor Spacing Distribution (NNSD)}

A nearest neighbor spacing $S_{1,k}$ for an unfolded eigenvalue $e_k$ is defined as  $|e_k-e_{\alpha}|$ with $e_{\alpha}$ as the nearest eigenvalue to $e_k$ on the complex plane: $S_{1,k}=min_{\alpha}|e_k-e_{\alpha}|$.  Due to ease of measurement, the spectral analysis through experimental as well as numerical route is  often based on the distribution $P(S, e)$ of unfolded nearest neighbor spacings $S$ at spectral point $e$ and thus a measure for short range correlations.  While $R_2(S; e)$ in general depends on  the eigenvalue spacings $S$ of all orders,  it  is approximately same as $P(S; e)$  for small $S$.  (Indeed, as in case of a real spectrum, ${\mathcal R}_2$ can  be expressed as a sum over $k^{th}$ order spacing distributions $P_k(S; e)$: 
\begin{eqnarray}
{\mathcal R}_2(S, e)=\sum_{k=0}^{\infty} P_k(S, e)
\label{r2s}
\end{eqnarray}
 with $P_0 \equiv P$ as the nearest neighbor spacing distribution \cite{fkpt5}).  Using the above relation in eq.(\ref{rna3}) then implies that the evolution of  $P_k(S, e)$ with changing system conditions and  from an arbitrary initial condition is governed by $\Lambda_e$ too for all $k$  including $k=0$.  

With changing system conditions,  the eigenvalues distribution  on a complex plane for many non-Hermitian systems undergoes a variation between two extremes, namely,  uncorrelated eigenvalues  (Poisson universality class) and correlated eigenvalues through mutual repulsion (Ginibre universality class). For the former,  $P(S; e)$ is the 2-dimensional Poisson distribution, independent of $e$: 
\begin{eqnarray}
P(S) =2 c_p S \; {\rm e}^{-\alpha_p \; S^2}
\label{pspoi}
\end{eqnarray}
with $c_p$ and $\alpha_p$ determined from the normalization condition $\int_0^{\infty} P(S) \; {\rm d}S=1$ and the unit mean level spacing condition $\int_0^{\infty} S \, P(S) \; {\rm d}S=1$. 
%
For the Ginibre ensemble of complex matrices,  the spacing distribution $P(S; e)$ (independent of $e$) is \cite{hkku}: 
\begin{eqnarray}
P(S) = c_g \; P_g(c_g S)
\label{psgin}
\end{eqnarray}
with 
\begin{eqnarray}
P_g(S) = \lim_{N \to \infty} \left[ \prod_{n=1}^{N-1} f_n(S^2) \; {\rm e}^{- S^2}\right] \; \left[ \sum_{n=1}^{N-1} \frac{2 S^{2n+1}}{n! \; f_n(S^2)} \right] 
\label{psgin0}
\end{eqnarray}
and 
\begin{eqnarray}
f_n(x) =\sum_{m=0}^{n}  \frac{x^m}{m!} = {\rm e}^{x} - \sum_{m=n+1}^{\infty}  \frac{x^m}{m!} 
\label{fx0}
\end{eqnarray}
and 
with $c_g$ as the mean level spacing $\int_0^{\infty} \; S \; P_g(S) \; {\rm d}S  \approx 1.1429$.  

From eq.(\ref{fx0}),  $f_n(x)$  for $S <1$  limit can be  approximated as $f_n(x) \sim 1$.  
This in turn gives,  for small-$S$,  $P(S) =(1/2) \; c_g^3 \; S^3 \; {\rm e}^{-c_g^2 \; S^2}$. For large $S ( >1)$,  $f_n(x)$  can be  approximated by its dominant term $f_n(x) \sim  \frac{x^n}{n!} $.  This in turn gives the dominant contribution  from the second bracket in eq.(\ref{psgin0}) as $S$ and from the first bracket as $\left(\prod_{n=1}^{N-1} \frac{S^{2n}}{n!}\right)  \; {\rm e}^{- S^2}$. 
For $S >1$, eq.(\ref{psgin0}) can then be approximated as $P(S) \sim S \; \left(\prod_{n=1}^{N-1} \frac{c_g ^{2n} \; S^{2n}}{n!}\right)  \; {\rm e}^{- c_g^2 \; S^2}$ or simply as $P(S) \sim \; {\rm e}^{- c_g^2 \; S^2}$ (with exponential term domunating over powers of $S$).

\section{Spacing Ratio Distribution} 

For a meaningful comparison of the statistics of two different ensembles or even those for a same ensemble at two different spectral locations,  it is necessary to  rescale the spectrum by  $R_{1, local}(e)$.  This however requires a prior knowledge of   $R_{1, local}(e)$ which is often not known for non-stationary ensembles and the standard route is to determine it through  numerical calculation.  Besides, in case  $R_1(e)$ is not a smooth function of energy,  the unfolding procedure becomes non-trivial even if $R_1(e)$ is analytically known and the spectrum is locally stationary.  Indeed as almost all standard spectral fluctuation measures e.g  nearest neighbour spacing distribution and number variance are sensitive to unfolding issues, 
%
it is preferable to consider a dimensionless measure e.g. the nearest neighbour spacing ratio. For real spectrum,  the measure  is defined as   \cite{huse}
 $P(r)=\sum_{i=1}^{N-1}\langle \delta(r-r_i)\rangle$ with  $r$ defined as the ratio of consecutive spacings between nearest neighbour levels: $r_i=S_{i+1}/S_i$ where $S_i=e_{i+1}-e_i$ is the distance between two  nearest neighbour eigenvalues. As the ratio $r$ does not depend on the local density of states,  an  unfolding of the spectrum for $P(r)$ is not required. Further $P(r)$ being a short range fluctuation measure,   the  probability of error due to mixing spectral 
 statistics is reduced.



\begin{figure}[ht!]
\centering

\vspace{-1.5in}

\includegraphics[width=20cm,height=30cm]{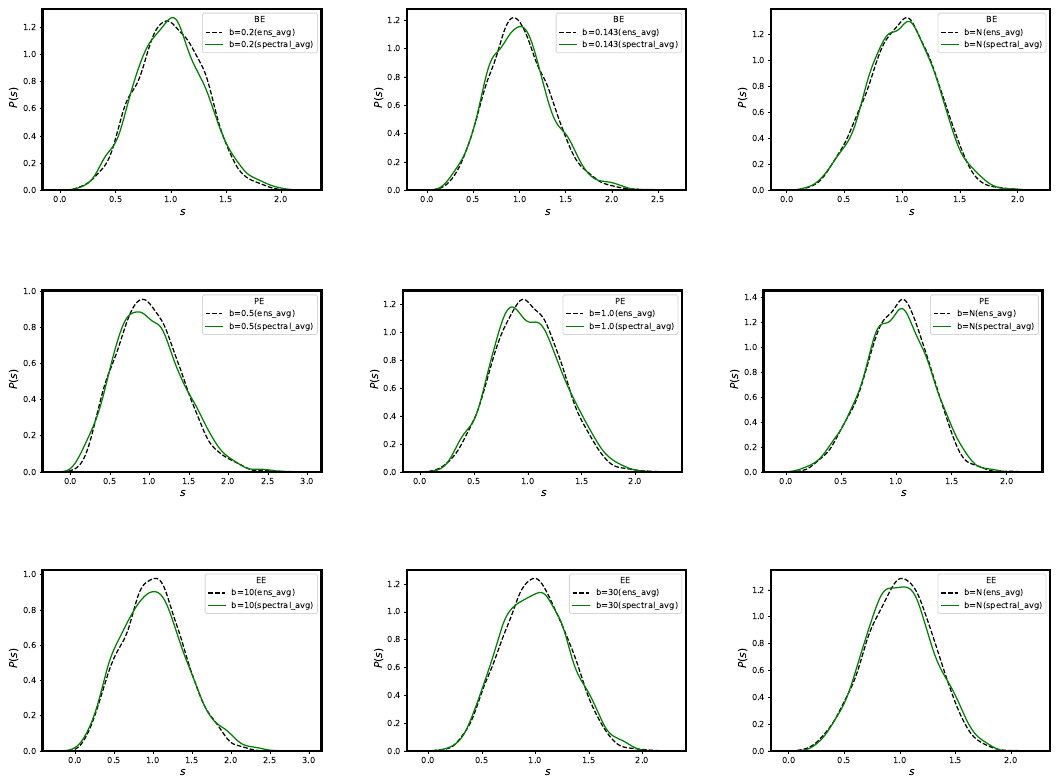}

\vspace{-4.0in}

\caption{{\bf Local ergodicity of the spectral fluctuations on the complex plane:} The figure displays a comparison  of  $P(S) \equiv P_e(S, e; \Lambda_e)$ obtained by ensemble averaging with that of spectral-ensemble averaging for three different $b$-values for each ensemble (i.e BE,  PE and EE).  
The ensemble averaging is obtained by choosing a single spacing (equivalently at a single spectral point) near $z \sim 0$ and averaging over an ensemble of $2500$  matrices each of size $N=1024$. The spectral-ensemble averaging is carried out by averaging over a few spacings ($\sim 100$) within a range $\Delta z$ around $z \sim 0$ from each matrix as well as over the ensemble of $25$ matrices; the total number of spacings is same for both the cases.    
As the visuals indicate, the two averages are almost  analogous.}
\label{nrbe}
\end{figure}

\begin{figure}[ht!]
\centering

\hspace{-3.5in}

\includegraphics[width=20cm,height=20cm]{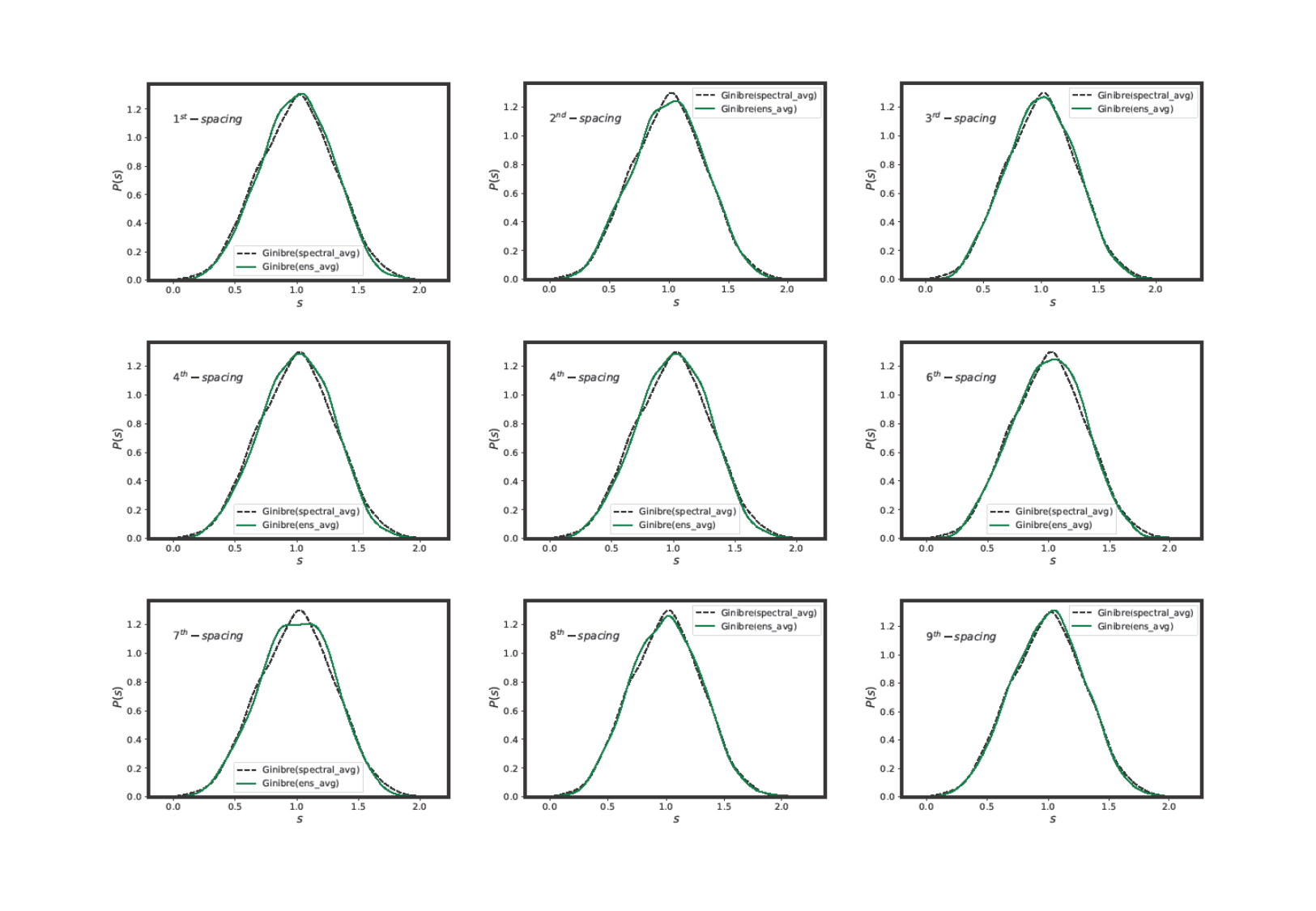}

\vspace{-0.5in}

\caption{{\bf Ergodicity of the spectral fluctuations on the complex plane for Ginibre ensemble :} The figure displays a comparison of  $P(S) \equiv P_e(S, e; \infty)$ obtained by ensemble averaging with that of spectral averaging,   for different choice of single spacings i.e different $e$ values.   The details of the two averaging  procedure are same as in figure 1. The two averages almost overlap in each case,  irrespective of the choice of the single spacing  used for the ensemble averaging.}
\label{nrpe}
\end{figure}

\begin{figure}[ht!]
\centering

\vspace{-0.9in}

\includegraphics[width=20cm,height=25cm]{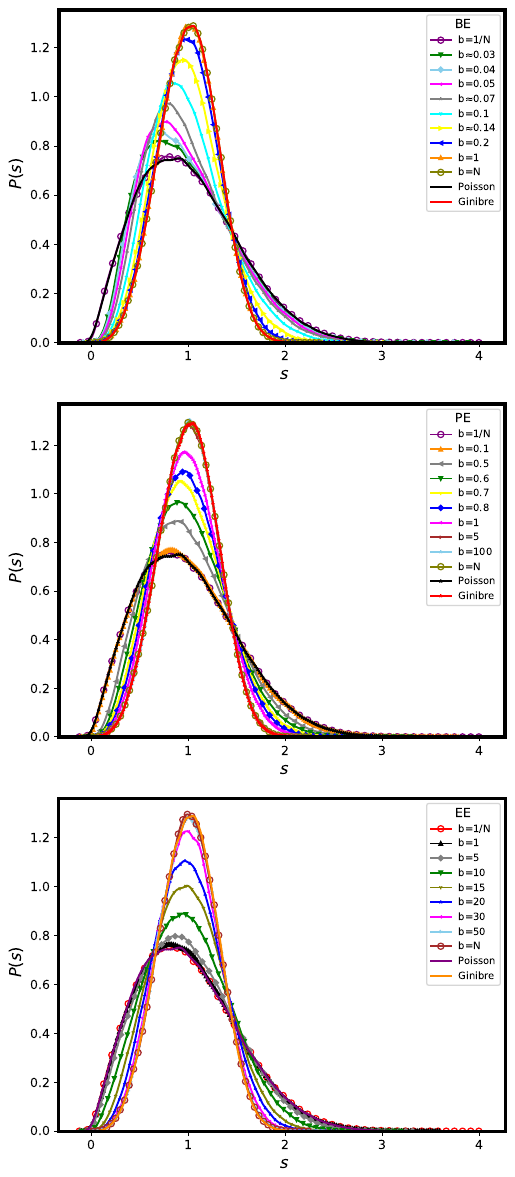}

\vspace{-1.8in}

\caption{{\bf  Single parameter governed evolution  of the fluctuation measures:}  The figure displays the  spectral-ensemble averaged nearest neighbour  spacing distribution $P(S) \equiv P_e(S, e; \Lambda_e)$ of the eigenvalues on the complex plane ($10 \%$ taken from the neighborhood for $e \sim 0$) for many $b$ values while $N$ is kept  fixed ($N=1024)$ for the three ensembles,  each consisting of $2500$ matrices.  As can be seen from the figure, a smooth  crossover from Poisson to Ginibre limit occurs for each case as $b$ and thereby $\Lambda_e$ varies.  A slight shift of the peak  towards left for BE cases  seems to be an artefact of the unfolding issues.   A good agreement of $P(S)$ for finite $\Lambda_e$ with eq.(\ref{psfit}) is displayed in figure \ref{psfit1} (the comparison displayed in a  separate figure for the purpose of clarity).}
\label{fps}
\end{figure}

\begin{figure}[ht!]
\centering

\vspace{-0.1in}

\includegraphics[width=20cm,height=30cm]{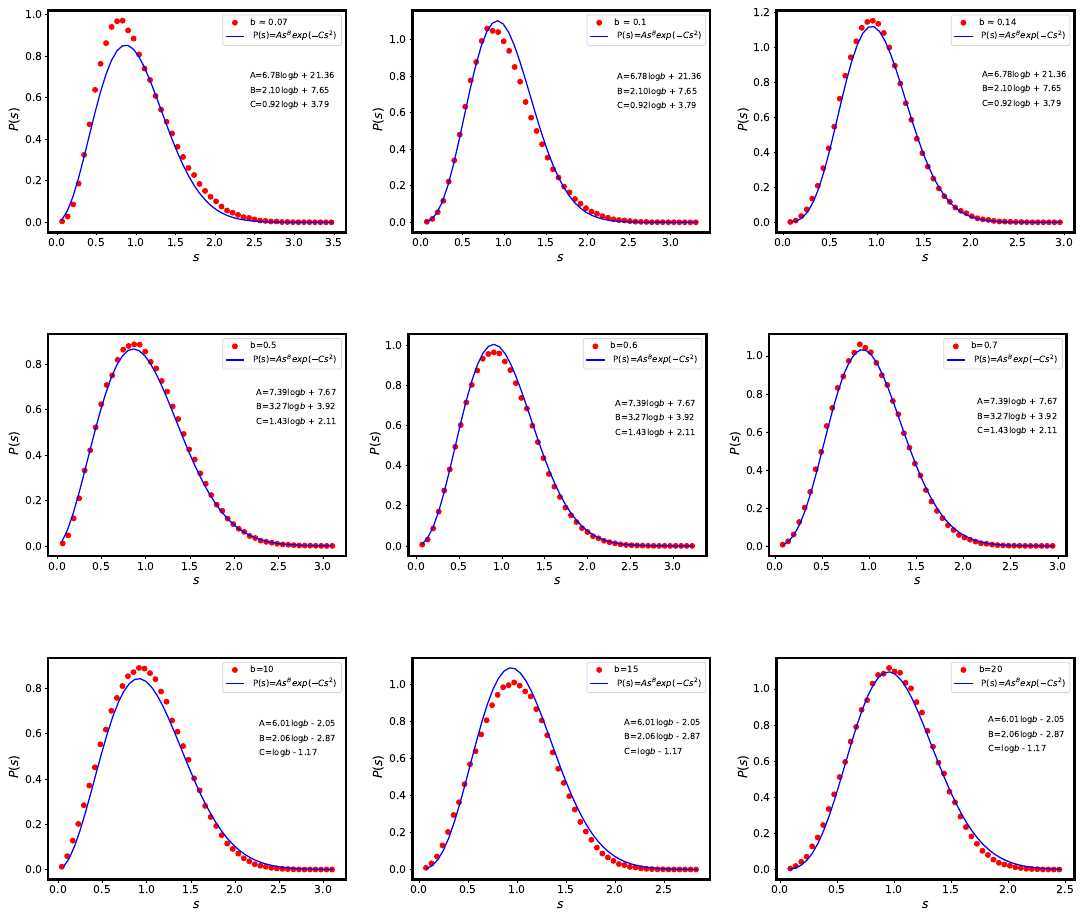}

\vspace{-4.2in}

\caption{{\bf  $P_e(S, e; \Lambda_e)$ for finite $\Lambda_e$: comparison with theoretical conjecture eq.(\ref{psfit}):}  The figure displays a comparison with  eq.(\ref{psfit}) of the $P(S) \equiv P_e(S, e; \Lambda_e)$  many $b$ values and a fixed $N=1024$ for the three ensembles, each consisting of $2500$ matrices.    As given in table 1,   the fitted parameters $A, B, C$ can be recast as the functions of $\Lambda_e$.  Here again the definition $\Lambda_e=\log b-C$ as the crossover parameter seemingly apply quite well.}
%
\label{psfit1}
\end{figure}

\begin{figure}[ht!]
\centering

\vspace{-0.9in}

\includegraphics[width=20cm,height=25cm]{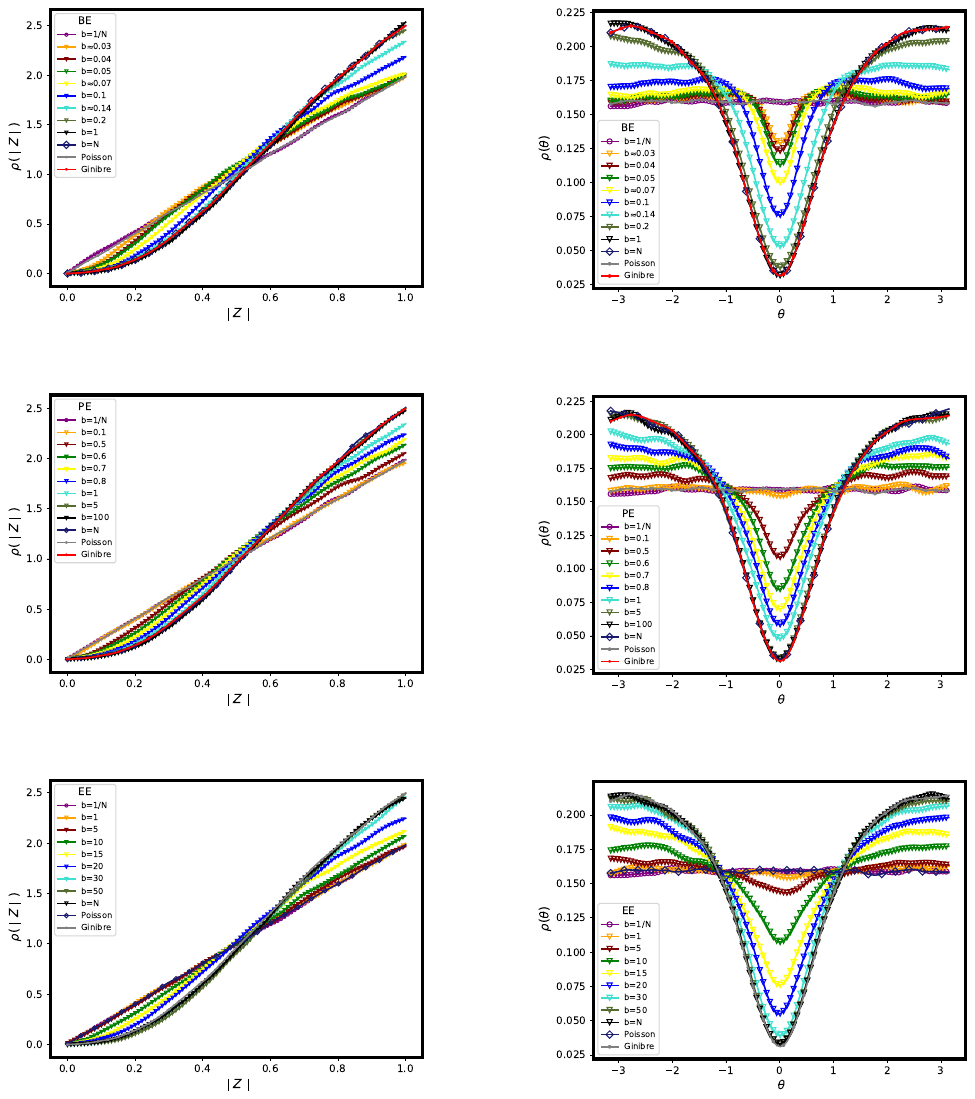}

\vspace{-2.3in}
\caption{{\bf  Radial and  angular dependence of spacing ratios:}  Figure illustrates the radial $\rho(|z|)$ and angular parts $\rho(\theta)$ of the  spectral-ensemble averaged nearest neighbour  spacing distribution $P_z(z)$ with $z=|z| {\rm e}^{i \theta}$ on the complex plane for many $b$ values for fixed system size $N=1024$ and ensembles size $M=5000$.   
As shown in left panels of  figure,  while $\rho(|z|)$  for intermediate $b$ values lies between Poisson and Ginibre limit in both regions $|z| < 0.5$ and $|z| > 0.5$, they seemingly converge to same point in the neighborhood of $|z| \sim 0.5$.  The display in right panel confirm the almost homogeneous $\rho(\theta)$ distribution in  Poisson limit for each ensemble but it rapidly changes with increasing $b$. Indeed $\rho(\theta)$ approaches a minimum for $\theta=0$ as $b$ approaches Ginibre limit.  This indicates that increasing level repulsion among consecutive eigenvalues with increasing $b$ causes them  to lie at large angular separations.}
\label{fps1}
\end{figure}

\begin{figure}[ht!]
\centering

\includegraphics[width=20cm,height=30cm]{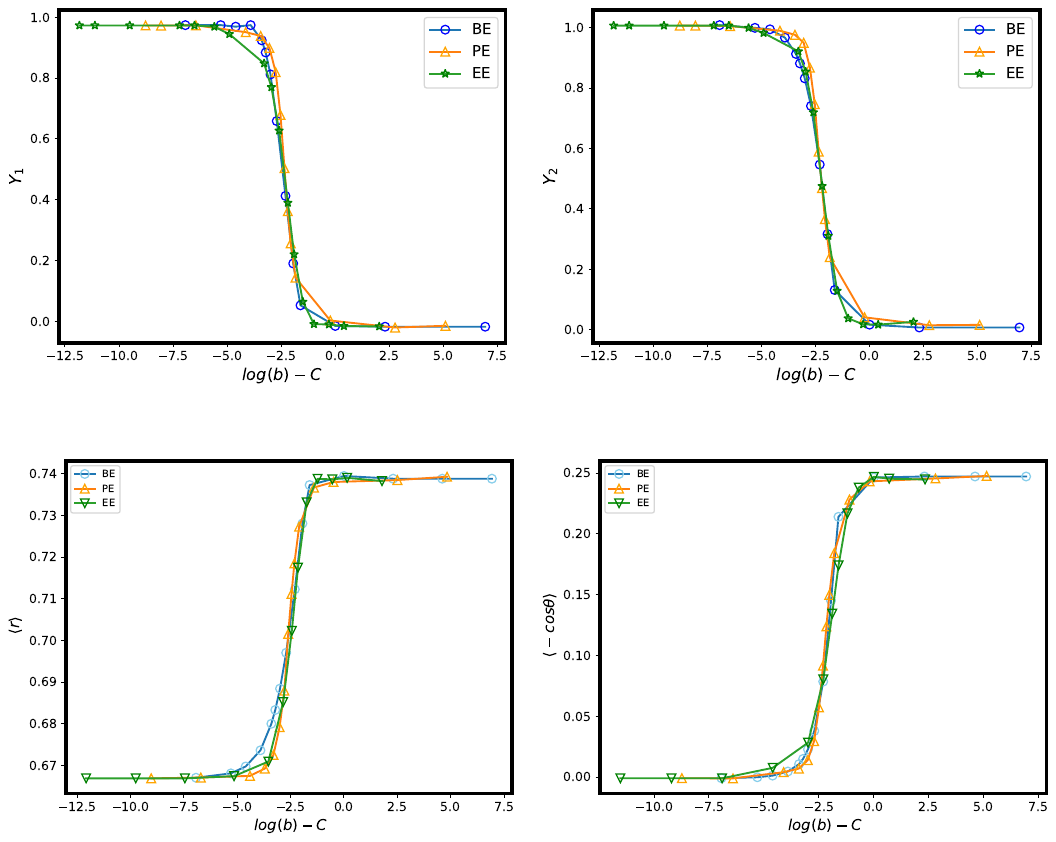}

\vspace{-4.8in}

\caption{{\bf  Universality of local fluctuations  in non-stationary non hermitian ensembles:}  The figure displays the behavior of four different fluctuation measures, namely,  (i) $\gamma(x_1)$,  (ii) $\gamma(x_2)$,  (iii) $\langle r \rangle$,  (iv) $-\langle \cos\theta \rangle $ 
on the complex spectral plane, with $x_1 =0.656978$ and $x_2=1.446259$. As can be seen from the figure, the behavior of each measure vs $\log b -c$  for three ensembles collapses onto same curve, with $c=0, 1.8, 4.6$ for BE, PE and EE respectively.  An important point worth noting is the rapid transition of each measure from one extreme limit to another,   indicating a critical statistics near $\log b -c \approx -2$.}
\label{fcps}
\end{figure}

\begin{figure}[ht!]
\centering

\vspace{-0.9in}

\includegraphics[width=20cm,height=25cm]{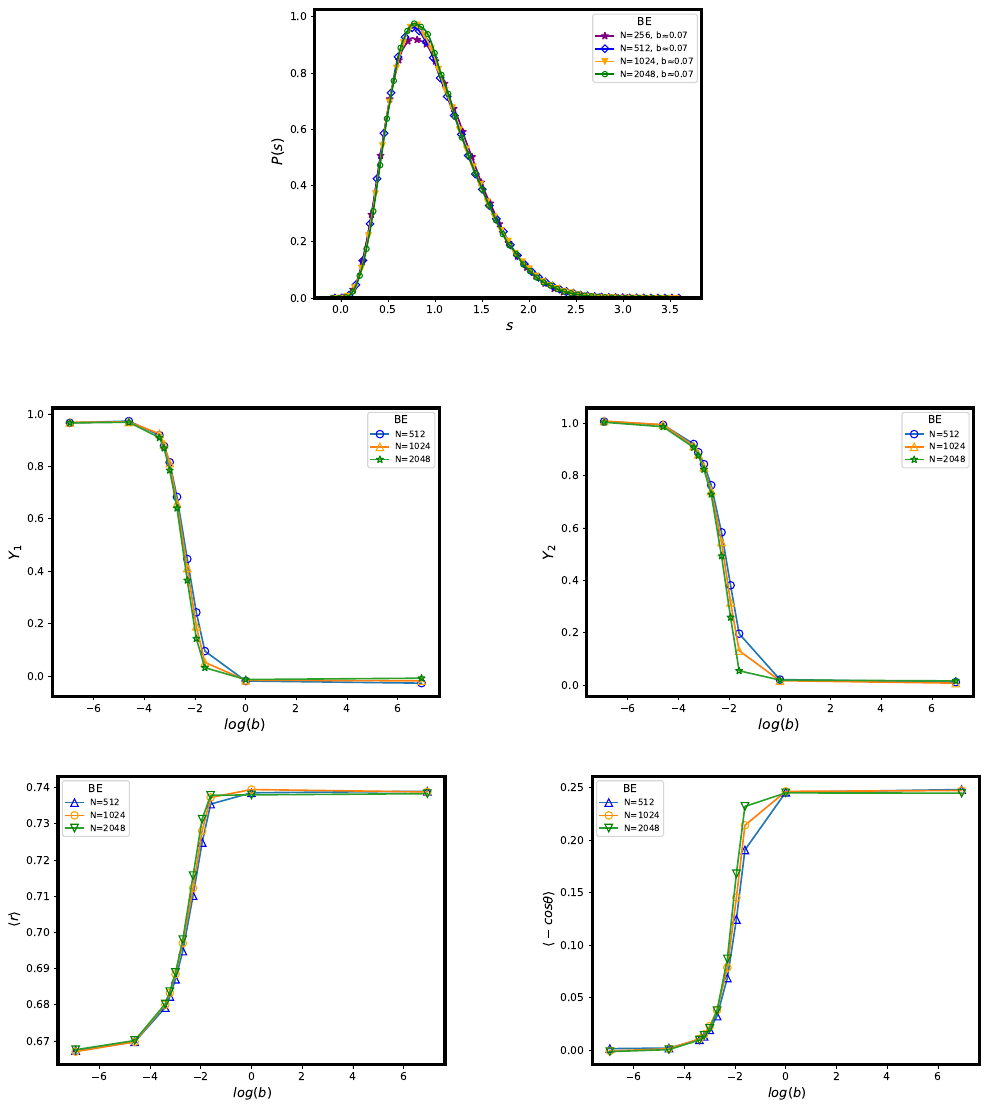}

\vspace{-1.8in}
\caption{{\bf  Critical BE Statistics} The figure displays the statistical behavior  for BE at  critical  values of the ensemble parameters (given in eq.(\ref{be1})) based on following fluctuation measures: (i) 
 $P(S)$, (ii) $\gamma_1$, (iii) $\gamma_2$,  (iv) $\langle r \rangle$, (v) $-\langle \cos \theta \rangle$.  As can be seen from each parts of the figure, the statistics for each measure remains invariant  with changing $N$ for entire range of $b$. This indicates the criticality of the ensemble for chosen set  of the ensemble parameters.  More clearly the spectral statistics near $e \sim 0$ is critical irrespective of the $b$ and $N$ values if the variances of the diagonal are taken in a specific combination described in eq.(\ref{be1}).  }
\label{critbe}
\end{figure}

\begin{figure}[ht!]
\centering

\vspace{-0.9in}

\includegraphics[width=20cm,height=25cm]{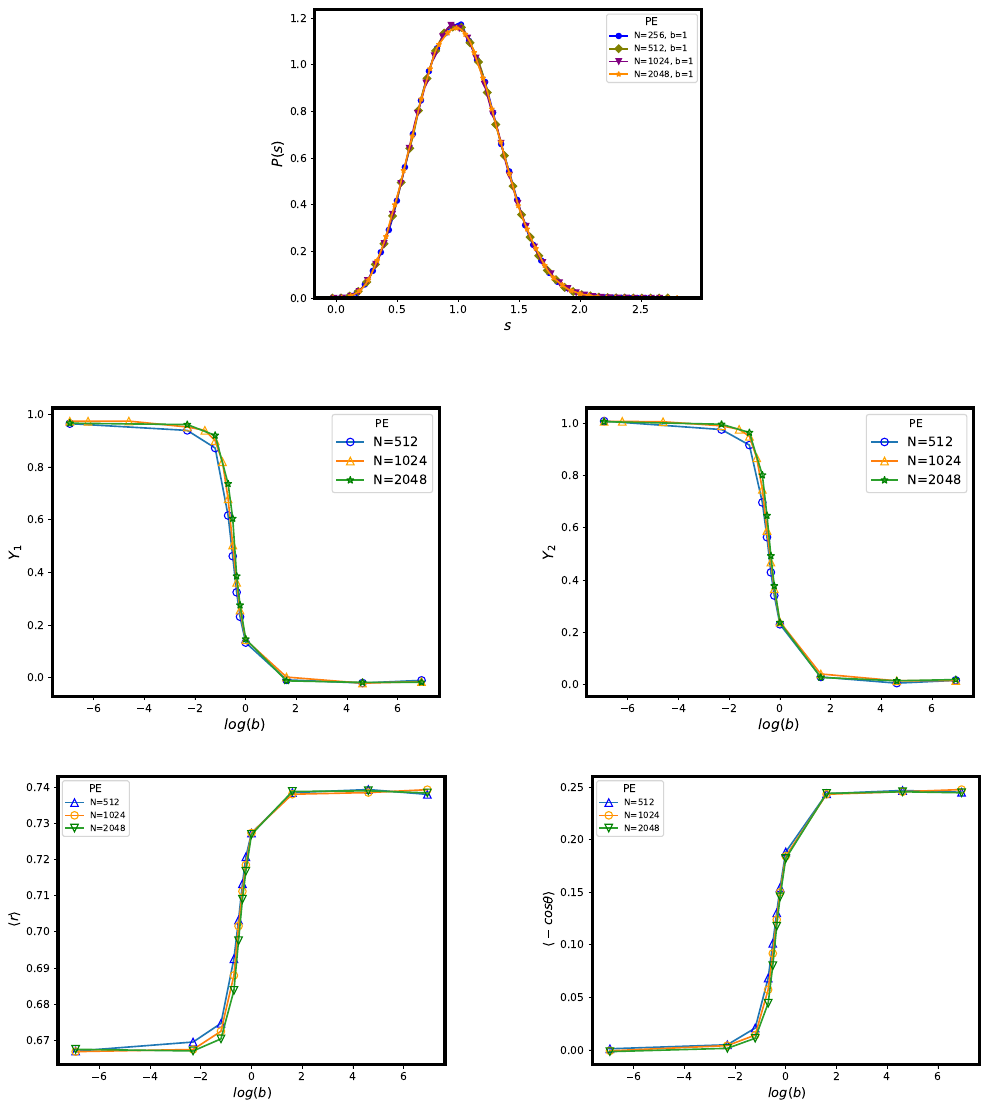}

\vspace{-1.8in}
\caption{{\bf  Critical PE Statistics:}  the details here are same as in figure 7 except now the critical PE ensemble is described by eq.(\ref{pe1}).}
\label{fps2}
\end{figure}

\begin{figure}[ht!]
\centering

\vspace{-0.9in}

\includegraphics[width=20cm,height=25cm]{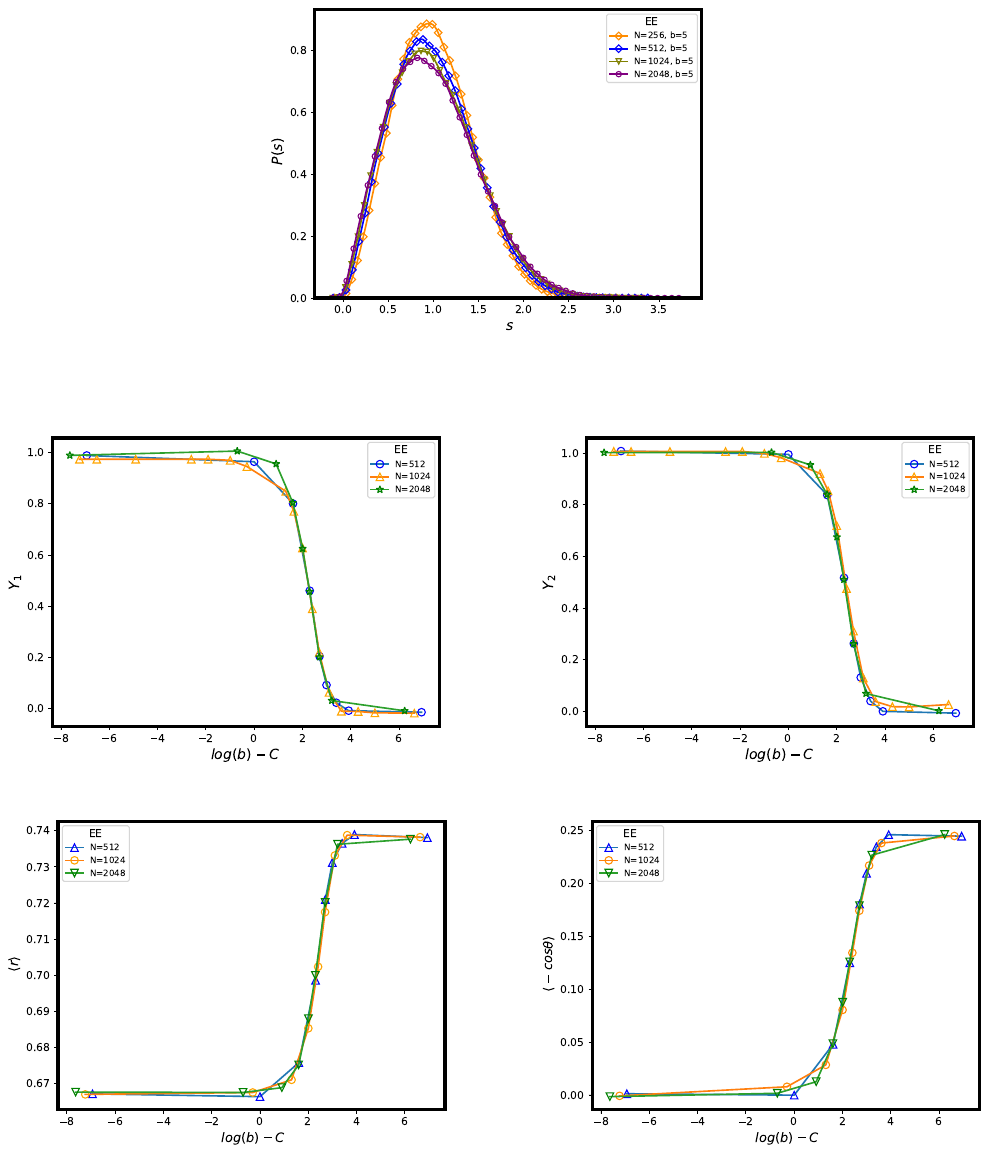}

\vspace{-1.8in}
\caption{{\bf  Critical EE Statistics:}  The details here are same as in figure 7 except now the critical EE ensemble is described by eq.(\ref{ee1}).}
\label{fps3}
\end{figure}

\end{document}